# Halide Perovskite-Chalcohalide Nanocrystal Heterostructures as a Platform for the Synthesis and Investigation of the CsPbCl$_3$-CsPbI$_3$ Epitaxial Interface

*Nikolaos Livakas, Irina Skvortsova, Juliette Zito, Yurii P. Ivanov, Aswin Asaithambi, Andrea Toma, Annick De Backer, Muhammad Imran, Sandra Van Aert, Giorgio Divitini, Ivan Infante\*, Sara Bals\* and Liberato Manna\**

N. Livakas, A. Asaithambi, L. Manna
Nanochemistry
Istituto Italiano di Tecnologia
Via Morego 30, Genova 16163, Italy
E-mail: liberato.manna@iit.it

Y. P. Ivanov, G. Divitini
Electron Spectroscopy and Nanoscopy
Istituto Italiano di Tecnologia
Via Morego 30, Genova 16163, Italy

A. Toma
Clean Room Facility
Istituto Italiano di Tecnologia
Via Morego 30, Genova 16163, Italy

I. Skvortsova, J. Zito, S. Van Aert, A. De Backer, S. Bals,
Electron Microscopy for Materials Science (EMAT)
University of Antwerp
2020 Antwerp, Belgium

I. Skvortsova, J. Zito, S. Van Aert, A. De Backer, S. Bals,
NANOlab Center of Excellence
University of Antwerp
2020 Antwerp, Belgium

I. Infante

BCMaterials

Basque Center for Materials, Applications, and Nanostructures

UPV/EHU Science Park

Leioa 48940, Spain

M. Imran,

Department of Electrical and Computer Engineering

University of Toronto

10 King's College Road, Toronto, ON M5S 3G4, Canada

Funding: Y.P.I. and G.D. acknowledge funding from the Italian Space Agency (Agenzia Spaziale Italiana, ASI) in the framework of the Research Day "Giornate della Ricerca Spaziale" initiative through the contract ASI N. 2023−4-U.0. N.L. and L.M. acknowledge funding from European Research Council through the ERC Advanced Grant NEHA (grant agreement n. 101095974). J.Z. and L.M. acknowledge funding from Ministero dell'Ambiente e della Sicurezza Energetica through the Project IEMAP (Italian Energy Materials Acceleration Platform) within the Italian Research Program ENEA-MASE (2021−2024 "Mission Innovation") (agreement 21A033302 GU no. 133/5−6−2021) as well as the extensive use of the CRESCO/ENEAGRID High Performance Computing infrastructure and the support of its management team. I.S. S.V.A., S.B. and L.M. acknowledge financial support from the Research Foundation-Flanders (FWO) through a doctoral fellowship (FWO-SB Grant No. 1SHA024N) and project funding (G0A7723N). A.T. acknowledges support by the European Research Council under the European Union's Horizon 2020 Research and Innovation Program through the ERC Consolidator Grant REPLY (grant agreement n. 101002422). S. B. and L. M. acknowledge funding from the European Union's Horizon Europe grant MSCA SE DELIGHT (GA n. 101131111).

**Abstract**: Halide exchange in lead-based halide perovskites has been studied extensively. While mixed Cl/Br or Br/I alloy compositions can be formed with no miscibility gaps, this is precluded to mixed Cl/I compositions, due to the large difference in $Cl^-$ and $I^-$ ionic radii. Here, we exploit perovskite-chalcohalide $CsPbCl_3$-$Pb_4S_3Cl_2$ heterostructures to study the Cl→I exchange and isolate new types of intermediate structures. The epitaxial interface between the $Pb_4S_3Cl_2$ chalcohalide and the $CsPbCl_3$ perovskite domain significantly influences the intermediate stages of halide exchange in the perovskite domain, leading to coexisting $CsPbCl_3$ and $CsPbI_3$ domains, thereby delivering segmented $CsPbI_3$-$CsPbCl_3$-$Pb_4S_3Cl_2$, energetically favorable heterostructures, with partial I-alloying of the $CsPbCl_3$ domain and at the perovskite-chalcohalide interface. The I:$CsPbCl_3$ domain between $CsPbI_3$ and $Pb_4S_3Cl_2$ enables a gradual lattice expansion across the heterostructure. This design accommodates interfacial strain, with a 5.6% mismatch at the $CsPbCl_3$–$CsPbI_3$ interface and a 3.4% mismatch at the perovskite–chalcohalide interface. Full halide exchange leads to $CsPbI_3$-$Pb_4S_3Cl_2$ heterostructures. Both in intermediate and fully exchanged heterostructures the $CsPbI_3$ domain is emissive. In the intermediate structures, the band alignment between the two perovskite domains is type-I, with the carriers photogenerated in the $CsPbCl_3$ domain quickly transferring to the $CsPbI_3$ domain, where they can recombine radiatively.

## 1. Introduction

Lead halide perovskite nanocrystals (NCs) have gathered significant attention due to their appealing optical properties and potential for various applications.[1–6] Soon after their synthesis was developed,[7,8] various studies addressed their reactivity.[9] In this context, halide exchange reactions are particularly important for these materials as they can be conveniently harnessed to fine-tune the optical band gap and achieve any desired emission color, from the blue to the red region of the visible spectrum.[10,11] In the past, almost all studies were focused on the Cl ↔ Br and Br ↔ I reactions, and it was assessed that all intermediate Cl-Br and Br-I alloy compositions can be prepared. On the other hand, the direct Cl ↔ I exchange was much less understood and studied, given the large difference in ionic radii between $Cl^-$ and $I^-$ ions.[10–12] Hence, it was unclear what types of intermediate structures could be formed during such exchange reactions. In a recent work of ours,[13] we explored the direct exchange of $Cl^-$ with $I^-$ ions in $CsPbCl_3$ NCs and discovered that the intermediate products of this reaction consist of a mixture of pristine (i.e. "unexchanged") $CsPbCl_3$ NCs and fully exchanged (i.e. $CsPbI_3$) NCs. Based on X-ray diffraction (XRD) analysis, both types of NCs had very limited alloy compositions (~6% $I^-$ in $CsPbCl_3$, and ~4% $Cl^-$ in $CsPbI_3$), at the edges of the miscibility gap. Neither NCs with $CsPb(Cl_{1-x}I_x)_3$ homogeneous alloy compositions nor with heteroepitaxial $CsPbI_3$-$CsPbCl_3$ interfaces could be isolated in those intermediate exchanged samples. The only alternative way for iodine to be added to the $CsPbCl_3$ NCs was to form a CsI surface layer.

While these results are in line with structural/thermodynamic predictions, our further aim was to investigate whether the introduction of additional structural constraints on a starting $CsPbCl_3$ NC could influence the outcome of such halide exchange reaction, especially at the intermediate stages. One possible structural constraint is represented by the epitaxial interface a $CsPbCl_3$ NC already shares with another material, as is the case for the perovskite-chalcohalide $CsPbCl_3$-$Pb_4S_3Cl_2$ NC heterostructures that were recently synthesized by us.[14,15] In the present work, we therefore investigated the Cl→I exchange reaction in such $CsPbCl_3$-$Pb_4S_3Cl_2$ NC heterostructures, and we were able to reveal distinct differences in the partial exchange case compared to free-standing $CsPbCl_3$ NCs. At intermediate exchange stages, the $CsPbCl_3$ domain farthest from the perovskite-chalcohalide interface undergoes complete anion exchange with $I^-$ ions, forming a heteroepitaxial $CsPbCl_3$-$CsPbI_3$ interface with the remaining portion of the $CsPbCl_3$ domain. Concurrently, $I^-$ ions penetrate and partially replace $Cl^-$ ions across three atomic layers at the perovskite-chalcohalide interface: the terminal CsCl monolayer of the perovskite domain converts to CsI, along with the shared atomic layer between the $CsPbCl_3$ and $Pb_4S_3Cl_2$ domains, and the first layer of the chalcohalide domain, forming a $Pb_4S_3(Cl_{1-x}I_x)_2$ structure. Interestingly, as the exchange progresses and the remaining $CsPbCl_3$ domain is reduced to a few atomic rows, it starts alloying with iodine, with $I^-$ ions preferentially settling in rows perpendicular to the $CsPbCl_3$-$CsPbI_3$ and perovskite-chalcohalide interfaces. The alloyed I:$CsPbCl_3$ domain, with a lattice parameter of 5.93 Å, which is intermediate between those of $CsPbI_3$ (6.26 Å) and $Pb_4S_3Cl_2$ (5.73 Å), enables a gradual lattice expansion that produces a graded structure, with ~5.6% lattice mismatch at the $CsPbCl_3$–$CsPbI_3$ interface and ~3.4% residual mismatch at the perovskite–chalcohalide interface. Density functional theory (DFT) calculations on fully atomistic models supported these observations. All these substitutions collectively lead to energetically favorable configurations during the various stages of the exchange process: first, initiating the Cl→I exchange at the bottom of the perovskite domain represents an energetic minimum among possible $Cl^-/I^-$ distributions, thus favoring the segmented $CsPbI_3$-$CsPbCl_3$-$Pb_4S_3Cl_2$ heterostructure. This energetic preference suggests that the region farthest from the interface, with fewer structural constraints, enables complete Cl→I exchange. Furthermore, DFT suggests that substituting $Cl^-$ with $I^-$ ions in the interfacial layers at the perovskite-chalcohalide interface lowers the total energy of the system. Likewise, the vertical ordering of $I^-$ ions in the remaining $CsPbCl_3$ domain at the later stage of exchange corresponds to an energetically favorable arrangement. Finally, provided enough $I^-$ ions are present, the exchange can be brought to completion, thus generating $CsPbI_3$-chalcohalide NCs. Remarkably, the partial (or full) exchange converts the initially non-emissive, type-I $CsPbCl_3$-$Pb_4S_3Cl_2$ heterostructures into emissive ones. In particular, the intermediate $CsPbI_3$-$CsPbCl_3$-$Pb_4S_3Cl_2$ heterostructure presents band alignments such that emission from the $CsPbCl_3$ domain is still quenched, while emission from the $CsPbI_3$ domain is possible. Transient absorption measurements evidenced a faster decay of the $CsPbCl_3$ exciton when $CsPbCl_3$ is interfaced to both $CsPbI_3$ and $Pb_4S_3Cl_2$ domains compared to the pristine $CsPbCl_3$-$Pb_4S_3Cl_2$ heterostructure, indicating the formation of a type-I alignment between the $CsPbCl_3$ and $CsPbI_3$ domains. In such configuration, the photoexcited carriers in the $CsPbCl_3$ can be transferred to either the chalcohalide or the $CsPbI_3$ domains when they are transferred to the $CsPbI_3$ domain they can recombine radiatively.

Overall, our work demonstrates that nanocrystal heterostructures enable the formation of epitaxial interfaces and alloy compositions (hence, new phases) that have not been reported so far neither in bulk solids nor in isolated nanocrystals.

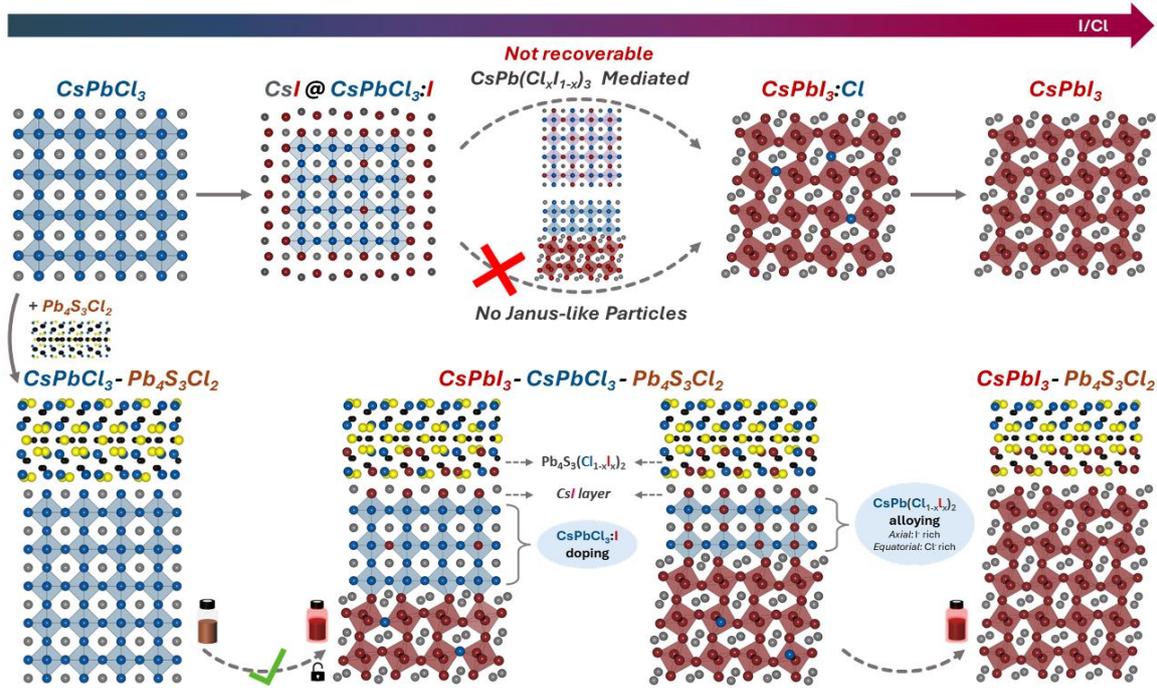

**Figure 1.** Schematic representation of the Cl→I exchange on both free-standing $CsPbCl_3$ NCs and $CsPbCl_3$-$Pb_4S_3Cl_2$ heterostructure NCs. As discussed in a previous work of ours, [13] anion exchange in the case of free-standing $CsPbCl_3$ NCs (top row of panels) proceeds through the initial formation of $CsPbCl_3$ NCs alloyed with ~5% iodine and covered with an exchanged CsI layer. As soon as the iodine concentration reaches its solubility limit in the $CsPbCl_3$ phase, the $CsPbCl_3$:I NCs rapidly turn into $CsPbI_3$:Cl NCs (with ~4% $Cl^-$ in $CsPbI_3$), jumping over the miscibility gap. Pure $CsPbI_3$ NCs are obtained after expelling the residual $Cl^-$ ions from the lattice. No other intermediate $CsPb(Cl_{1-x}I_x)_3$ alloy compositions nor phase-segregated $CsPbI_3$-$CsPbCl_3$ intermediates could be recovered.[13] In the case of the $CsPbCl_3$-$Pb_4S_3Cl_2$ NC heterostructures of this work, the same reaction follows a different pathway (bottom row of panels). The exchange starts with $I^-$ ions replacing $Cl^-$ at the perovskite-chalcohalide interface and at the bottom of the $CsPbCl_3$ domain, converting part of it into $CsPbI_3$ and forming an epitaxial $CsPbCl_3$-$CsPbI_3$ interface. As the reaction progresses, most of the perovskite domain transforms into $CsPbI_3$, reducing the $CsPbCl_3$ domain to a few atomic rows in projection. At this late stage of the exchange, the $CsPbCl_3$ domain exhibits considerable alloying with $I^-$ ions. Also, these ions are preferentially distributed along the vertical direction. This process effectively leads to segmented $CsPbI_3$-$CsPbCl_3$-$Pb_4S_3Cl_2$ red-emitting heterostructures. With enough $I^-$, fully exchanged $CsPbI_3$-chalcohalide heterostructure NCs are obtained.

## 2. Results and Discussion

The CsPbCl$_3$/Pb$_4$S$_3$Cl$_2$ NC heterostructures studied in this work were prepared following the procedure published by our group in a previous work.[14] Briefly, the synthesis consists of a two-step approach: the first step involves the preparation of CsPbCl$_3$ clusters at a relatively low temperature (50 °C) over a long reaction time (30 min). The second step involves consecutive injections of Pb-oleate, dodecanethiol, CsPbCl$_3$ clusters, and elemental sulfur (dissolved in octadecene) into pre-degassed octadecene at 200 °C under inert atmosphere (see Experimental Section). The prepared CsPbCl$_3$/Pb$_4$S$_3$Cl$_2$ NCs were then subjected to I$^-$ exchange by employing PbI$_2$ (in the presence of oleylamine and oleic acid) as a halide source. **Figure 2** reports optical absorption (ABS) and photoluminescence (PL) spectra, along with X-ray diffraction (XRD) patterns of the initial and I$^-$ exchanged samples. The optical spectra (**Figure 2a**) and the XRD patterns (**Figure 2b**) demonstrate the progressive, partial to complete halide exchange, depending on the amount of PbI$_2$ precursor added. Consistent with our prior findings on isolated CsPbCl$_3$ NCs,[13] PbI$_2$ incorporation induces the formation of the CsPbI$_3$ phase alongside the initial CsPbCl$_3$ phase (and the Pb$_4$S$_3$Cl$_2$ here). This new CsPbI$_3$ phase is responsible for the PL emission in the red spectral region, while no emission is seen in the initial heterostructures. With increasing iodine source, the PL peak red shifts until the reaction reaches the stage of fully exchanged CsPbI$_3$/Pb$_4$S$_3$Cl$_2$ NCs (**Figure 2a,b**).

High-angle annular dark-field scanning transmission electron microscopy (HAADF-STEM) analysis indicates that halide exchange does not alter the overall NCs' morphology (**Figure 2c-e**). However, energy dispersive X-ray spectroscopy maps acquired in STEM mode (STEM-EDX, see **Figure 2f-h** and **Figure S1**) reveal compositional changes. Elemental analysis of a single partially exchanged NC evidence of a significant presence of iodine alongside chlorine in the perovskite domain, suggesting the coexistence of CsPbCl$_3$ and CsPbI$_3$ phases within the same particle (**Figure 2g** and **Figure S1**). Iodine species are also detected at the perovskite-chalcohalide interface and surrounding the chalcohalide domain. Interestingly, HAADF-STEM images from NC that appear to be in the early stages of the reaction indicate the formation of a CsI passivation layer on the CsPbCl$_3$ surface (**Figure S2**), similarly to what was observed in isolated CsPbCl$_3$ NCs.[13] The iodine incorporation is also observed at the perovskite–chalcohalide interface (**Figure S2, red arrows**). In fully exchanged CsPbI$_3$/Pb$_4$S$_3$Cl$_2$ NCs, iodine remains present around the chalcohalide domain (**Figure 2h**). Notably, the chalcohalide domain never gets exchanged beyond its surface region, indicating a very low diffusivity of the halide ions in such material.

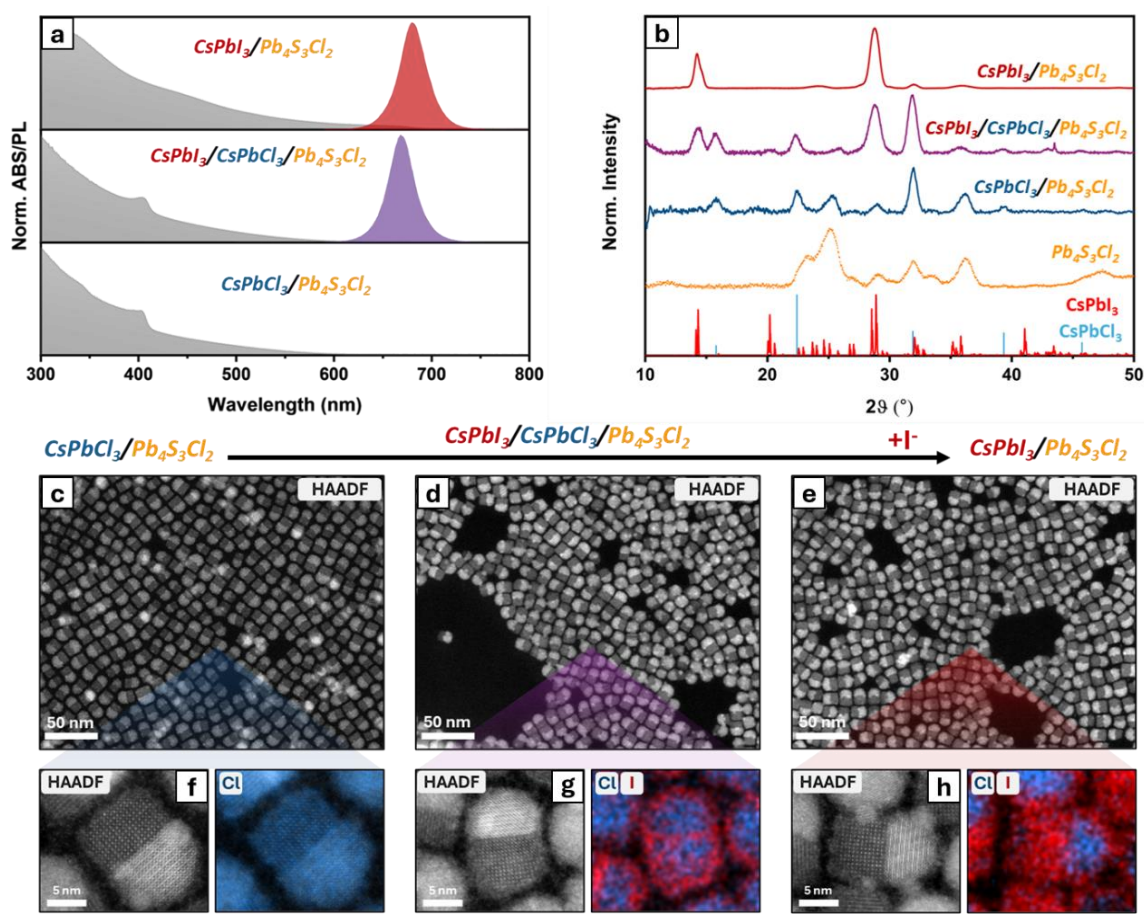

**Figure 2.** Steady-state optical, XRD, and electron microscopy analyses of the initial, intermediate and final samples for the Cl→I exchange reaction on $CsPbCl_3$-$Pb_4S_3Cl_2$ NC heterostructures. (a) Optical absorption (grey) and PL (colored) spectra of the pristine, partially exchanged ($X_{cl}$=52%) and fully exchanged ($X_{cl}$=5%) heterostructures ($X_{cl}$, defined as: $X_{cl}$ = [Cl]/([Cl]+[I])*100, is calculated from STEM-EDX data). XRD patterns (b) and HAADF-STEM images (c-e) along with the corresponding STEM-EDX elemental maps (f-h). In (b), the XRD pattern of a powder sample of separately synthesized $Pb_4S_3Cl_2$ NCs is also shown for comparison.

HAADF-STEM imaging typically requires relatively high electron doses to achieve a sufficient signal-to-noise ratio. It is, however, well-known that these doses can induce beam damage and ion diffusion within the perovskite structure, leading for example to the formation of Pb clusters on the NC surface.[16–18] To overcome this challenge and to further investigate the Cl→I exchange on heterostructures, we employed 4D-STEM with an event-driven direct electron detector. In earlier works, a convolutional neural network (CNN) was trained to output the phase image from such datasets.[19,20] A main advantage of this this approach is that the required electron dose could be reduced from >1000 e$^-$/Å$^2$, as in case of HAADF-STEM, to 300 e$^-$/Å$^2$, which is important for the investigation of local changes in structure and composition without the influence of the electron beam. As illustrated in **Figure 3a-c**, there are only minor changes in the shape of the exchanged heterostructures in comparison to the initial CsPbCl$_3$/Pb$_4$S$_3$Cl$_2$ NCs. Specifically, in partially exchanged NCs, the top region of the perovskite domain, connected to the chalcohalide domain, remains compressed in the direction parallel to the interface (**Figure 3b**). This morphological feature might indicate the presence of structurally different regions within the perovskite domain. To assess this hypothesis, we performed octahedral tilt measurements based on the CNN images (**Figure 3d-f**, see details in the experimental section). As expected, the octahedral tilt for the $\alpha$-CsPbCl$_3$ domain in the initial heterostructure is close to zero (**Figure 3d**), whereas for the fully exchanged $\gamma$-CsPbI$_3$ domain in the final exchanged heterostructure octahedral tilt values reach up to 14 degrees (**Figure 3f**). Interestingly, for the partially exchanged case (**Figure 3e**), we can distinguish two different areas with tilt distributions fluctuating around 0 and 14 degrees, which effectively demonstrates the coexistence of CsPbCl$_3$-CsPbI$_3$ domains sharing an epitaxial interface. Such distribution of phases in partially exchanged NCs reveals unique Cl→I exchange dynamics in the heterostructures, different from the free-standing NCs observed by us in our previous work.[13]

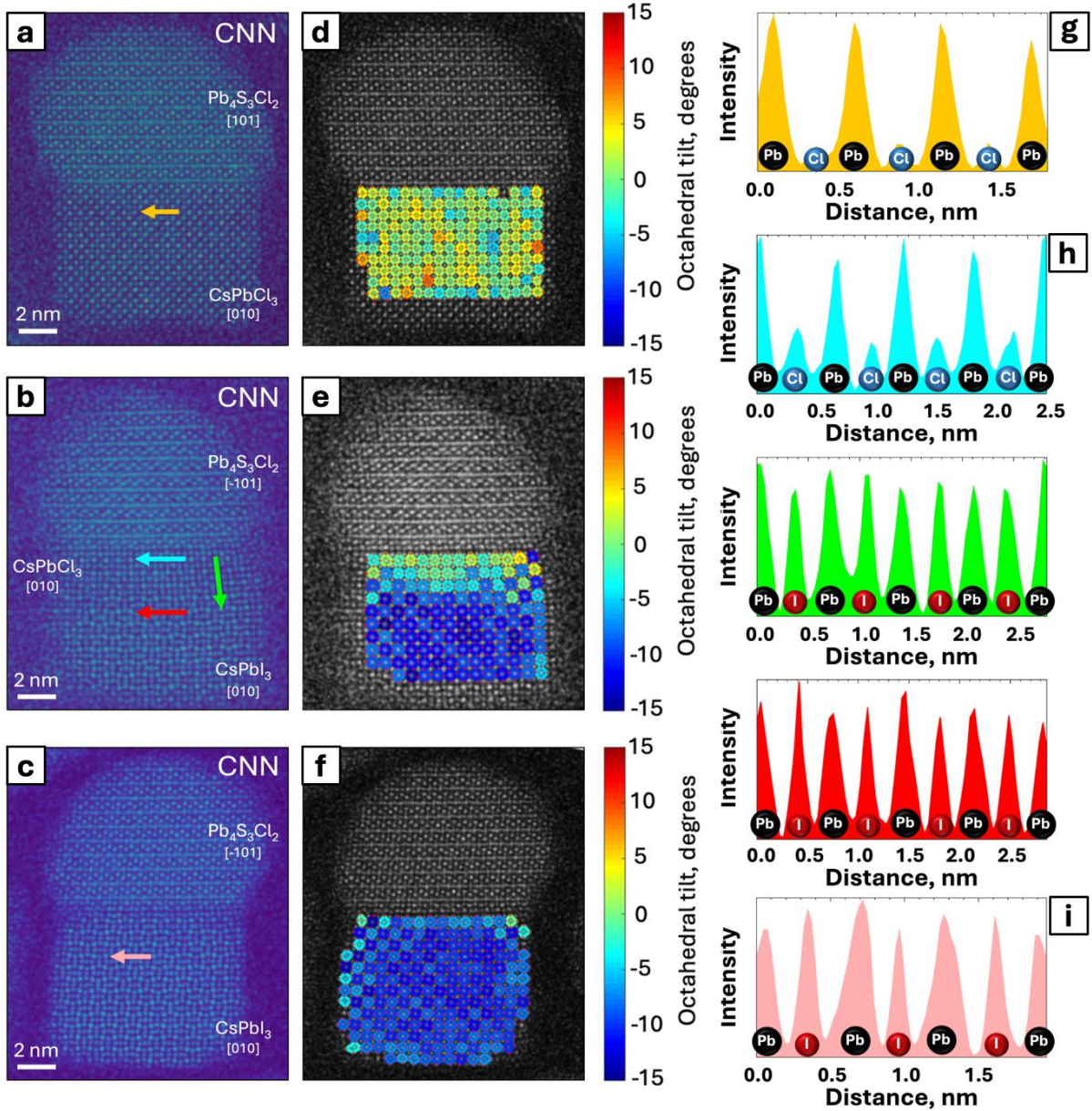

**Figure 3.** Phase images obtained from CNN reconstruction of 4D-STEM dataset of the starting $CsPbCl_3$-$Pb_4S_3Cl_2$ (a), partially exchanged $CsPbCl_3$-$CsPbI_3$-$Pb_4S_3Cl_2$ (b) and fully exchanged $CsPbI_3$-$Pb_4S_3Cl_2$ (c) heterostructures. The color-coded octahedral tilt mapping (d,e,f) calculated for (a,b,c). Intensity profiles (g,h,i) correspond to the arrows in (a,b,c) where Pb+X columns are labelled as Pb and pure X columns are labelled as Cl or I. Although some degree of iodine may be present in chlorine columns and chlorine in iodine columns, the labels denote the predominant element in each column. The octahedral tilt analysis estimates the averaged angle between the lattice vectors and the vectors corresponding to the direction of $PbX_6$ octahedra, thus, reflecting Pb-Pb-X angle.

Notably, in the CNN-reconstructed images, the intensities of the atomic columns are more sensitive to light elements than in HAADF-STEM images.[20] Thus, this approach enables direct visualization of both chlorine and iodine atomic columns (**Figure 3a,c**), such that they can be reliably distinguished (**Figure 3g,i**). Therefore, in addition to structural information, the CNN-reconstructed images provide qualitative insight into the relative composition of halide atomic columns based on their intensity variations. In the following lines, we focus on a single heterostructure at a relatively advanced stage of exchange as this reveals a wide richness of features. From the intensity profiles (**Figure 3h**, **Figure S3**, and **Figure S4**) derived from the CNN image of this heterostructure we can draw several conclusions. The profiles confirm the coexistence of $CsPbCl_3$-$CsPbI_3$ domains, as illustrated by the red and cyan intensity profiles in **Figure 3h** as well as additional supporting profiles, provided in **Figure S3**. As seen in the cyan profile in **Figure 3h**, the intensity peaks located between the Pb+X columns, corresponding to pure halide columns, are substantially lower than those observed in the red profile. The peak heights in the cyan profile resemble those of Cl columns (**Figure 3g**), whereas the red profile aligns more closely with the intensity of I columns (**Figure 3i**). Interestingly, the $CsPbCl_3$ domain appears to be alloyed with iodine, with the $I^-$ ions concentrated primarily along the vertical direction within the $CsPbCl_3$ lattice (green profile in **Figure 3h** and **Figure S4**). In contrast, we also observed NC heterostructures with slightly larger $CsPbCl_3$ domains (evidently captured at an early stage of halide exchange) showing no evidence of iodine incorporation, as confirmed by intensity profiles in both vertical and horizontal directions (**Figure S5**). The comparison between these two heterostructures (**Figure 3b and Figure S5**) provides insight into the exchange dynamics: initially, distinct domains of pure $CsPbCl_3$ and $CsPbI_3$ within a single heterostructure share an epitaxial interface. As the halide exchange progresses, this is followed by a reduction in size of the $CsPbCl_3$ domain and by its substantial alloying with iodide ions, the latter replacing chloride ions predominantly along directions vertical to the $CsPbCl_3$-$CsPbI_3$ interface, with the cubic crystal structure in the $CsPbCl_3$ region being preserved.

The lattice parameter analysis for the partially exchanged heterostructures (**Figure S6**) reveals lattice expansion when moving away from the perovskite-chalcohalide interface, supporting the structural distinction between the $CsPbCl_3$ (cubic phase, *Pm-3m*) and $CsPbI_3$ (gamma phase, *Pnma*) domains, as shown earlier with the octahedral tilt measurements (**Figure 3e**). A similar intensity and structural investigation was performed on the partially exchanged heterostructures where the $CsPbI_3$ domain is aligned along the [101] direction (**Figure S7**), and the results are in good agreement with the conclusions presented above, proving the presence of two distinct structural subdomains (**Figure S7d,e** and **Figure S8**) within the perovskite region as well as significant alloying of the CsPbCl$_3$ domain with iodine (**Figure S7b**).

Intensity profile analyses of CNN images also offer qualitative insights concerning the composition near the perovskite-chalcohalide interface (**Figure 4**). A CsI layer separates the perovskite and chalcohalide domains (**Figure 4c**). Such iodine segregation at the perovskite-chalcohalide interface was also observed in the STEM-EDX maps of **Figure 2g,h**. The atomic layers in the chalcohalide domain immediately adjacent to the interface are also alloyed with iodine, thus leading to a Pb$_4$S$_3$(Cl$_{1-x}$I$_x$)$_2$ composition there (**Figure 4d,e**). Further beyond these layers, the peak heights of the halide columns decrease (**Figure 4f**), suggesting that deeper regions in the chalcohalide domain preserve their pristine Pb$_4$S$_3$Cl$_2$ composition.

To rationalize the difference in behavior of heterostructures versus free-standing CsPbCl$_3$ nanocubes upon halide exchange we further investigated the epitaxial relation between chalcohalide and perovskite phases. Although CsPbCl$_3$ crystallizes in the cubic *Pm-3m* space group and Pb$_4$S$_3$Cl$_2$ adopts an orthorhombic *Pnma* structure, their lattice parameters are sufficiently close to allow epitaxial matching. On the other hand, CsPbI$_3$ crystallizes in the *Pnma* space group, the same as Pb$_4$S$_3$Cl$_2$. The lattice mismatch was calculated based on the average lattice parameters measured from HAADF-STEM images (**Figure S9**) for each phase through the following expression:

$$\text{lattice mismatch} = \left( \frac{a_{perovskite} - a_{chalcohalide}}{a_{perovskite}} \cdot 100\% \right)$$

For the initial CsPbCl$_3$–Pb$_4$S$_3$Cl$_2$ heterostructures, the lattice mismatch is almost negligible (≈0.4%, **Figure S9a,b,c**). In the case of partially exchanged heterostructures, the presence of an alloyed I:CsPbCl$_3$ domain between CsPbI$_3$ and Pb$_4$S$_3$Cl$_2$ allows a gradual expansion of the lattice parameters (**Figure S9d,e,f**). The mismatch at the CsPbCl$_3$–CsPbI$_3$ interface is ≈5.6%, whereas the mismatch at the perovskite–chalcohalide interface is ≈3.4% (**Figure S9e,f**). However, this latter value can be expected to already include some compensation due to iodide incorporation into the chalcohalide region close to the interface (**Figure 4d,e** and **Figure S9f**), which has a Pb$_4$S$_3$(Cl$_{1-x}$I$_x$)$_2$ composition. The observed overall gradient in lattice mismatch most likely favors the formation of CsPbI$_3$-CsPbCl$_3$-Pb$_4$S$_3$Cl$_2$ heterostructures and makes it possible to capture the CsPbI$_3$-CsPbCl$_3$ as intermediates of the halide exchange reaction in our experiments. . After complete halide exchange, the mismatch at the perovskite-chalcohalide interface is ≈8.5% (**Figure S9g,h,i**). Again, this value, obtained for the final heterostructures, includes compensation due to the presence of a Pb$_4$S$_3$(Cl$_{1-x}$I$_x$)$_2$ layer in the chalcohalide region close to the interface. Experimentally, the interfacial atomic distances parallel to the perovskite-chalcohalide interface are measured to be 5.69 ± 0.08 Å for the initial heterostructures (**Figure S9c, black box**) and 5.92 ± 0.18 Å for the fully exchanged ones (**Figure S9i, black box**). These values, extracted from the spacing of the single atomic row at the interface, indicate that the perovskite lattice is compressed, whereas the chalcohalide lattice is expanded in order to maintain the epitaxial relation.

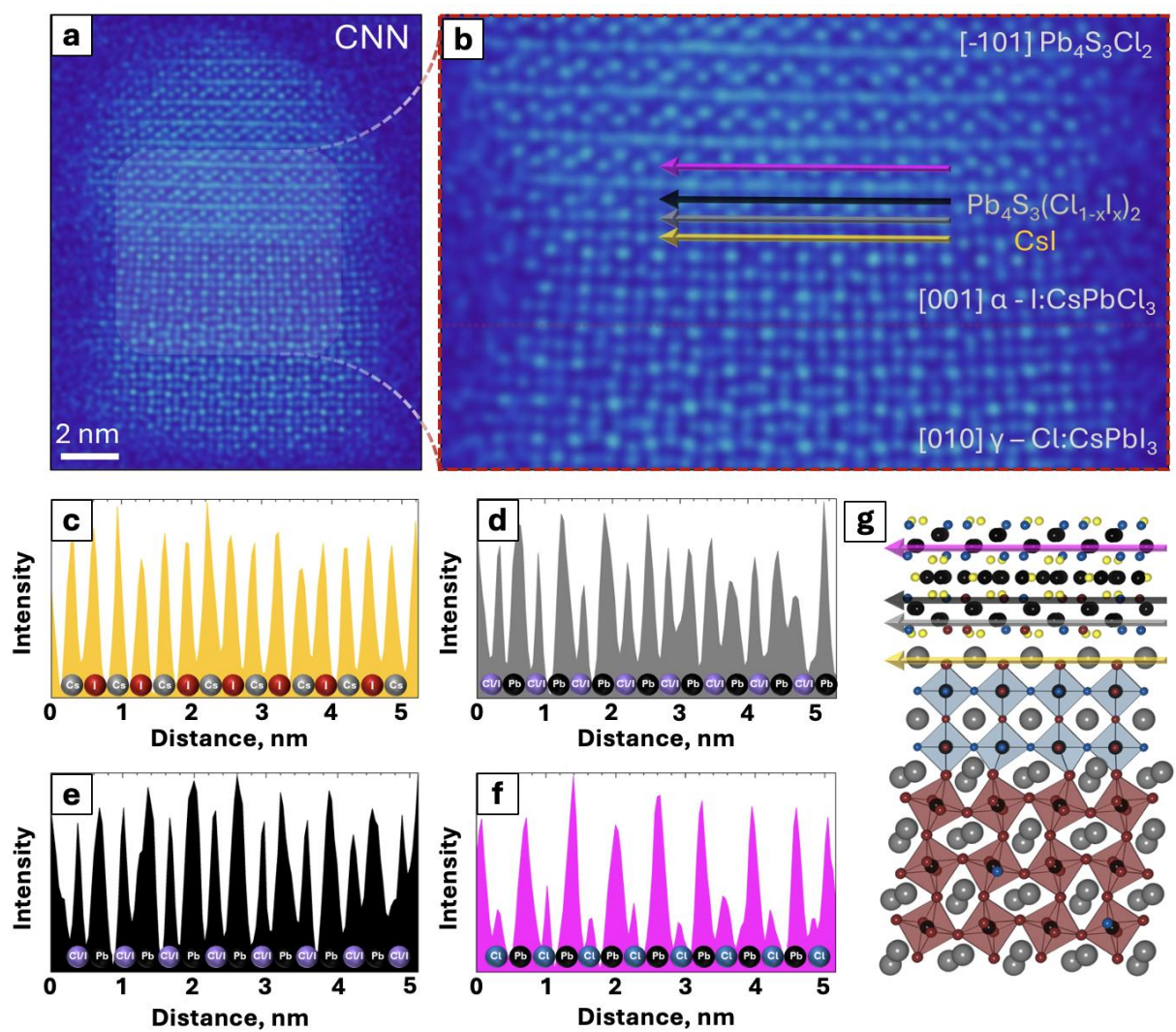

**Figure 4.** (a) Phase image obtained from CNN reconstruction of a 4D-STEM dataset of the partially exchanged $CsPbI_3$-$CsPbCl_3$-$Pb_4S_3Cl_2$ NC heterostructure presented in **Figure 3b**. (b) Enlarged area from the same high-resolution CNN image. The arrows correspond to the intensity profiles represented in panels (c-f) showing three atomic layers at the $CsPbCl_3$-$Pb_4S_3Cl_2$ interface exchanged with $I^-$ (c,d,e). (g) Reconstructed model corresponding to the multidomain structure presented in panel (a).

To shed light onto the formation and stability of these intermediate multidomain structures, we performed a series of DFT simulations. Here, we mimic the experimental conditions by including PbI$_2$ in the reaction equilibrium with the Cl-based heterostructure and we assume for simplicity that the displaced chlorine is released as PbCl$_2$. Because of this, all Cl→I anion exchange reactions were described by the following exchange mechanism:

$$HS(x \cdot Cl) + x\, PbI_2 \rightarrow HS(x \cdot I) + x\, PbCl_2$$

where $HS(x \cdot Cl)$ and $HS(x \cdot I)$ are the NC heterostructure models before and after the exchange reaction, and $x$ is the number of exchanged halides. The associated reaction energies were computed as:

$$\Delta E_{exchange} = [(E_{HS(x \cdot I)} + E_{PbCl_2}) - (E_{HS(x \cdot Cl)} + E_{PbI_2})]/x$$

that, they are normalized to the number of exchanged halide ions. We additionally computed the lattice strain on the perovskite subdomain from the Pb–Pb distances, which are directly related to the lattice vectors of the cubic perovskite structure. The strain was evaluated by comparing the Pb–Pb distances in the heterostructures to those in the corresponding standalone perovskite NC, according to the expression:

$$\text{strain (\%)} = \frac{d_{Pb-Pb}^{HS} - d_{Pb-Pb}^{perovskite}}{d_{Pb-Pb}^{perovskite}} \cdot 100$$

In this framework, the interfacial strain is defined as the average absolute value of the strain on the Pb–Pb distances within the interfacial perovskite layer. We started from the CsPbCl$_3$-Pb$_4$S$_3$Cl$_2$ NC heterostructure model published previously by us (**Figure 5a**)[14] 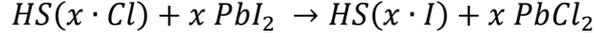 and gradually replaced Cl$^-$ ions with I$^-$ ions in five layers of the perovskite domain, either starting from the CsPbCl$_3$-Pb$_4$S$_3$Cl$_2$ interface (**Figure 5b**) or from the bottom of the perovskite domain (**Figure 5c**). A glance at energies associated with the Cl→I exchange in each layer, reported in **Table S1**, highlights that the partial substitution of Cl$^-$ with I$^-$ has a slight preference for the bottom of the perovskite domain, favoring the formation of a segmented CsPbI$_3$-CsPbCl$_3$-Pb$_4$S$_3$Cl$_2$ NC heterostructure, as observed in the experiments. This pathway is not only energetically favored but also maintains a low interfacial strain on the perovskite (1.57% for **Figure 5c**), comparable to the unexchanged heterostructure (1.66%), whereas the alternative pathway starting from the interface results in a higher strain (2.04% for **Figure 5b**).

We then investigated the structural and energetic changes associated with the partial substitution of Cl$^-$ with I$^-$ at the interface between the perovskite and chalcohalide domains. We started from the CsPbI$_3$-CsPbCl$_3$-Pb$_4$S$_3$Cl$_2$ NC heterostructure model of **Figure 5c** and fully replaced Cl$^-$ with I$^-$ in the interfacial layer of the perovskite domain (single monolayer). This is represented in **Figure 5d** and essentially reproduces the interfacial CsI layer observed in the CNN image of **Figure 4b,c**. We then gradually replaced Cl$^-$ ions with I$^-$ ions in the first two layers of the chalcohalide domain, achieving in these layers I/Cl ratios of 20%, 40%, 60%, 80%, and 100%, respectively (**Figure 5e**). As shown by the energy values reported in **Table S2**, the initial introduction of I$^-$ ions into the interfacial chalcohalide layers (from 0% to 20%) is the most energetically favored process, an effect that we ascribe to a substantial relaxation of the interfacial strain. Thereafter, the enthalpy change remains essentially constant as the I/Cl ratio increases. We note that entropic contributions, which are not included in our calculations, may play a critical role in stabilizing intermediate I/Cl ratios at around 40-60% (**Figure 5e**), as the number of possible configurations for halide ion distribution is maximized in this range. This mixed distribution is in line with the intensities of the halide columns observed in the CNN images.

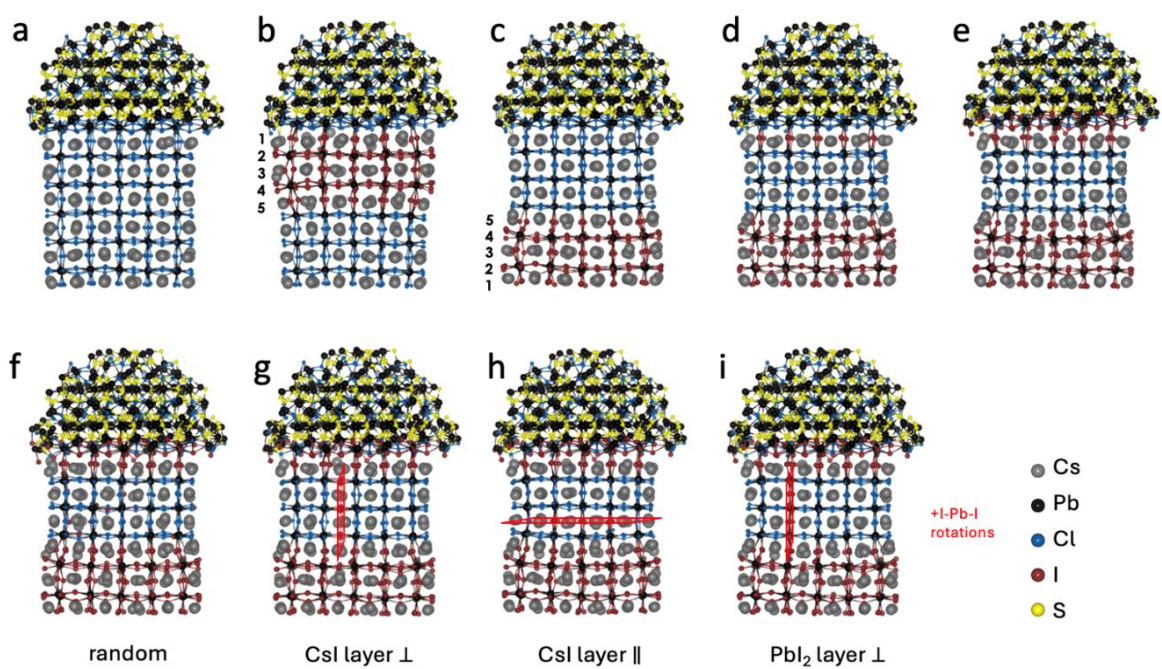

**Figure 5.** Ball and stick representation of heterostructure models optimized at the DFT/PBE level of theory (Cs$^+$ = grey, Pb$^{2+}$ = black, Cl$^-$ = blue, I$^-$ = red, S$^{2-}$ = yellow). (a) pristine CsPbCl$_3$-Pb$_4$S$_3$Cl$_2$; (b, c) anion exchanged samples in five perovskite layers, respectively starting from the CsPbCl$_3$-Pb$_4$S$_3$Cl$_2$ interface or from the bottom of the perovskite domain, the latter being slightly favored energetically. For model (c), further anion exchange reactions were modelled at the interface, including (d) full exchange of a CsCl monolayer in the perovskite domain, and (e) additional partial exchange of about 60% of Cl with I in two layers of the chalcohalide

domain, reducing the interfacial strain. (f-i) Model (d) after exchanging about 20% of the Cl$^-$ ions with I$^-$ ions in the CsPbCl$_3$ domain. Four possible distributions of the I$^-$ ions were probed: (f) randomly in the CsPbCl$_3$ domain; ordered (g) in a CsI layer perpendicular to the interface, (h) in a CsI layer parallel to the interface and (i) in a PbI$_2$ layer perpendicular to the interface. The ordered configurations are always preferred over the random one.

Finally, we simulated the formation of an I:CsPbCl$_3$ alloy by substituting about 20% of the Cl$^-$ ions with I$^-$ ions in the CsPbCl$_3$ domain. Guided by the electron microscopy analysis (Figure 3), we decided to probe four possible distributions of the I$^-$ ions: randomly in the CsPbCl$_3$ domain (**Figure 5f**); ordered in a CsI layer perpendicular to the interface (**Figure 5g**), in a CsI layer parallel to the interface (**Figure 5h**) and in a PbI$_2$ layer perpendicular to the interface (**Figure 5i**). As reported in **Table S3**, the formation of ordered I:CsPbCl$_3$ configurations is always preferred, with a stabilization energy of about 5-15 kcal/mol over the random configuration, which corresponds to a stabilization of 0.33-1.00 kcal/mol per atom being exchanged. The total stabilization energy is expected to be amplified in the NC heterostructures of experimental size, where a larger number of halide ions are exchanged. A closer look at the relaxed structure of **Figure 5g,i** reveals how ordered arrangements of I$^-$ ions can result in the expansion of the CsPbCl$_3$ lattice parallel to the interface (due to the elongation of the X-Pb-X bonds), ultimately decreasing the strain at the CsPbCl$_3$-CsPbI$_3$ interface. These results help us rationalize the ordered arrangement of I$^-$ ions in perpendicular layers within the CsPbCl$_3$ domain observed in the electron microscopy images (green arrow in Figure 3b).

We then computed the electronic structure of models of the starting CsPbCl$_3$-Pb$_4$S$_3$Cl$_2$ (**Figure 5a**)[14], the partially exchanged CsPbI$_3$-CsPbCl$_3$-Pb$_4$S$_3$Cl$_2$ (**Figure 5e**) and the fully exchanged CsPbI$_3$-Pb$_4$S$_3$Cl$_2$ heterostructures. The results are reported in **Figure 6a-c**. In all the cases, the valence band (VB) and conduction band (CB) edges of the heterostructure are defined by surface trap states on the chalcohalide domain (grey lines in the density of states, molecular orbital plots (3) and (4)), a result that is consistent with our previous work on the CsPbCl$_3$-Pb$_4$S$_3$Cl$_2$ heterostructures.[14] For CsPbCl$_3$-Pb$_4$S$_3$Cl$_2$, the energetic alignment between the bands of Cl-based perovskite domain and the bands of the chalcohalide domain gives rise to a mixed perovskite-chalcohalide configurations in both CB and VB (dashed ovals in Figure 6a). The driving force provided by the energy level alignment of these traps, together with the orbital overlap between the two domains, is expected to lead to efficient transfer of photoexcited electrons and holes to chalcohalide trap states. This enables non-radiative recombination pathways, explaining the lack of emission from the starting CsPbCl$_3$-Pb$_4$S$_3$Cl$_2$ heterostructure. In contrast, for the fully exchanged CsPbI$_3$-Pb$_4$S$_3$Cl$_2$ heterostructure the band misalignment results in VB states localized on the CsPbI$_3$ domain, while CB states remain mixed (dashed ovals in **Figure 6c**). Here, photo-excited holes are more likely to radiatively recombine with the electrons in the perovskite domain, accounting for the observed emission from the partially exchanged CsPbI$_3$-CsPbCl$_3$-Pb$_4$S$_3$Cl$_2$ and fully exchanged CsPbI$_3$-Pb$_4$S$_3$Cl$_2$ heterostructures. Additionally, in the partially exchanged case, the presence of the CsPbCl$_3$ domain prevents (both physically and electronically) the transfer of photo-excited electron and holes from the CsPbI$_3$ to the chalcohalide states, further facilitating the radiative recombination of the charge carriers in the I-based perovskite domain. We note that these considerations are based on ground-state DFT and should therefore be regarded as qualitative trends; more rigorous insight into excited-state recombination dynamics would require TDDFT or related approaches, which remain computationally prohibitive for nanocrystal models of this size.

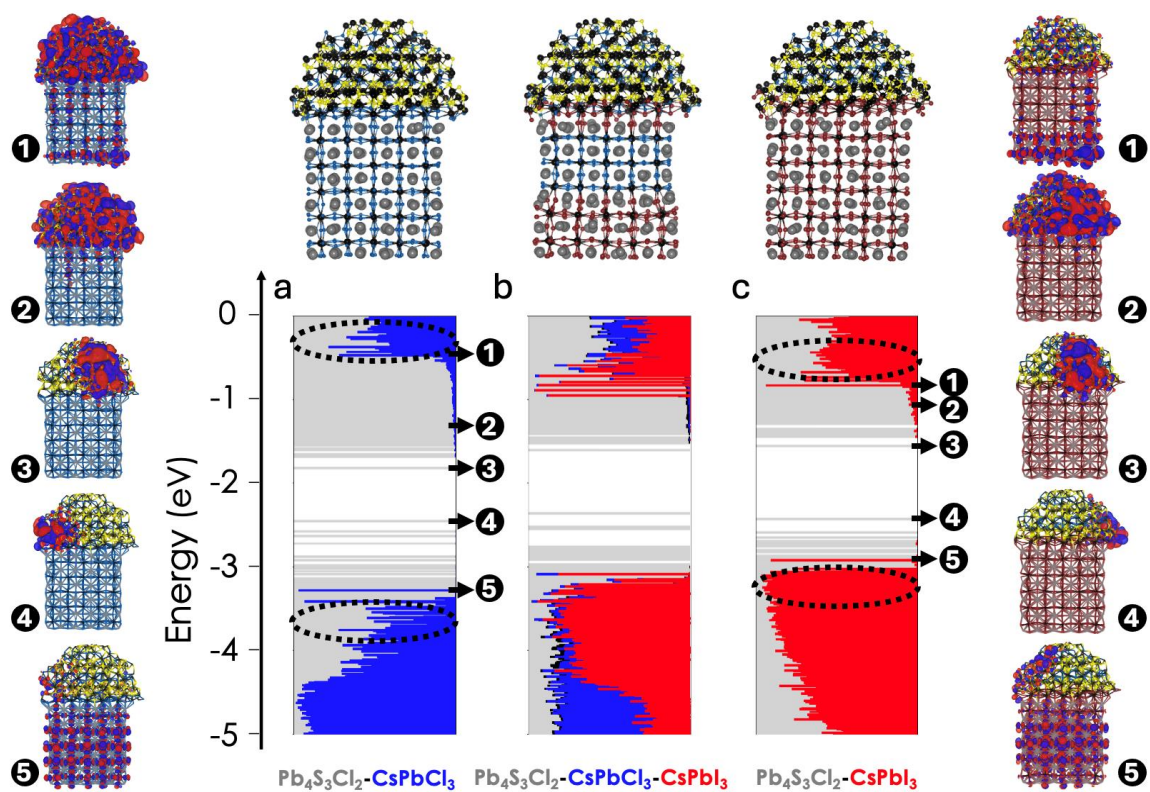

**Figure 6.** The electronic structure of (a) the CsPbCl$_3$-Pb$_4$S$_3$Cl$_2$, (b) the CsPbI$_3$-CsPbCl$_3$-Pb$_4$S$_3$Cl$_2$, and (c) the CsPbI$_3$-Pb$_4$S$_3$Cl$_2$ heterostructures computed at the DFT/PBE level of theory. For each molecular orbital (MO), the length of the different line sections represents the fractional contribution from the chalcohalide (grey), CsPbCl$_3$ (blue), and CsPbI$_3$ (red) domains,

respectively. Isosurfaces of the most relevant molecular orbitals are additionally reported for the parent and fully exchanged heterostructures with a counter value of 0.02 e/Bohr$^3$, highlighting positive and negative parts in red and blue, respectively. MO1 and MO5 highlight delocalized perovskite contributions in both CB and VB; MO2 illustrates delocalized chalcohalide contributions in the CB; MO3 and MO4 represent the CB and VB edges of the heterostructures, defined by trap states localized at the surface of the chalcohalide domain.

We carried out transient absorption (TA) spectroscopy on the starting, intermediate and fully exchanged heterostructures, as well on free-standing CsPbCl$_3$ and CsPbI$_3$ NCs for comparison. The study focused on ground-state bleaching (GSB) of CsPbCl$_3$ and CsPbI$_3$ excitons as key indicators of carrier behavior. Two different wavelengths of the laser pump were systematically employed: 370 nm, which is above the bandgap of both CsPbCl$_3$ and CsPbI$_3$ domains, and 500 nm, which selectively excites only that of CsPbI$_3$. **Figure 7a** presents the TA spectra at a 1 ps delay for CsPbCl$_3$ NCs, initial CsPbCl$_3$-Pb$_4$S$_3$Cl$_2$ and intermediate CsPbI$_3$-CsPbCl$_3$-Pb$_4$S$_3$Cl$_2$ heterostructures. In agreement with the steady-state absorption measurements (**Figures 1a, S10**), the GSB signal of the CsPbCl$_3$ exciton is observed around 405 nm in the heterostructures and 408 nm for the CsPbCl$_3$ NCs. Furthermore, positive features on both the higher- and lower-energy sides of the exciton GSB are attributed to photoinduced absorption (PA) processes and biexciton signals, respectively.[21–24] **Figure 7b** shows the decay profiles of exciton GSB for the CsPbCl$_3$ domain in each sample. The reference free-standing NCs (cyan trace) exhibit a significantly slower decay rate, with long exciton recombination times (nanosecond timescale). In contrast, the CsPbCl$_3$ exciton GSB in the starting CsPbCl$_3$-Pb$_4$S$_3$Cl$_2$ heterostructure decays more rapidly (blue trace), an effect that can be attributed to the formation of a type-I alignment between these domains.[14] A faster GSB decay compared to the free-standing case would indeed result from the transfer of carriers from the CsPbCl$_3$ domain to the Pb$_4$S$_3$Cl$_2$ one upon photoexcitation. The decay dynamics of the CsPbCl$_3$ exciton GSB in the pristine CsPbCl$_3$-Pb$_4$S$_3$Cl$_2$ heterostructures suggests a multiexponential behavior, where the fast components are related to charge transfer to the Pb$_4$S$_3$Cl$_2$ domain and the slow component can be assigned to exciton recombination in the CsPbCl$_3$ region. Upon partial exchange and the formation of a CsPbI$_3$ domain in the heterostructure, the CsPbCl$_3$ exciton GSB decays at an even faster rate (purple line) suggesting that most of the excited charge carriers in the CsPbCl$_3$ domain are quickly transferred across the CsPbCl$_3$-CsPbI$_3$ interface.

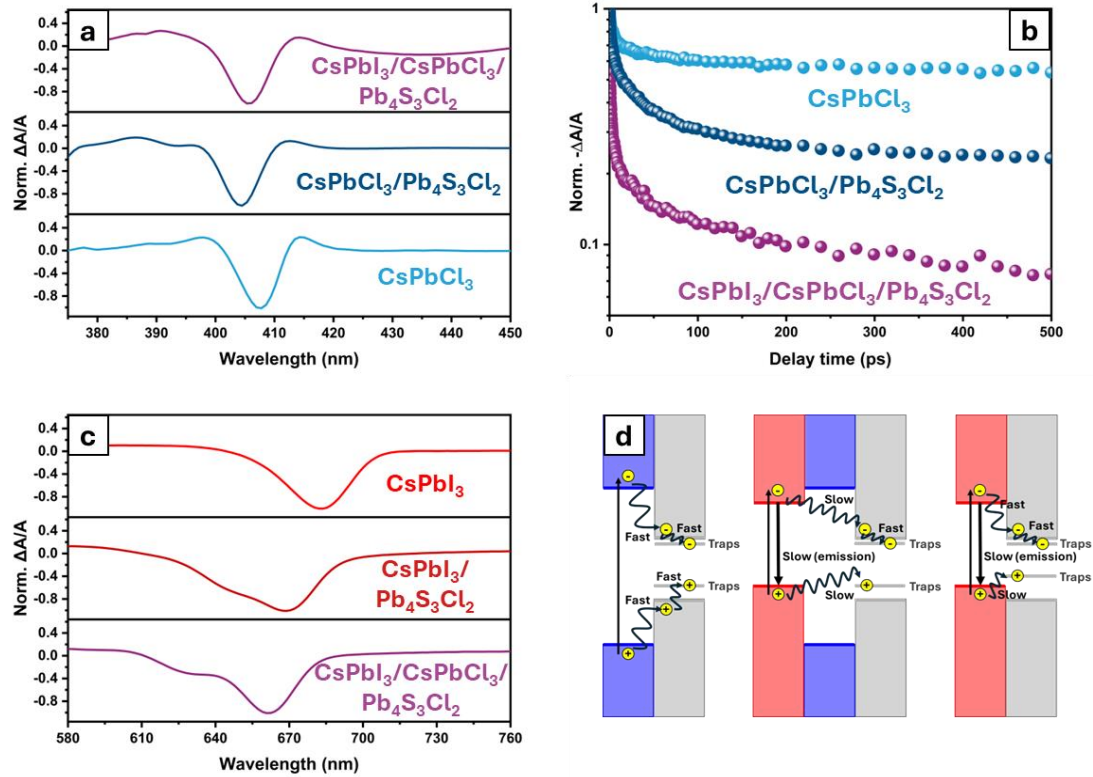

**Figure 7.** Transient absorption spectra at 1 ps delay for CsPbCl$_3$, CsPbCl$_3$-Pb$_4$S$_3$Cl$_2$, and CsPbI$_3$-CsPbCl$_3$-Pb$_4$S$_3$Cl$_2$ heterostructures excited at 370 nm. (b) Decay profiles of CsPbCl$_3$ exciton GSB in each sample. c) Transient absorption spectra at 1 ps delay for the CsPbI$_3$, CsPbI$_3$-Pb$_4$S$_3$Cl$_2$, and CsPbI$_3$-CsPbCl$_3$-Pb$_4$S$_3$Cl$_2$ heterostructures. (d) Schematic band-diagram of the heterostructures (CsPbCl$_3$ in light blue; Pb$_4$S$_3$Cl$_2$ in grey; CsPbI$_3$ in light red) illustrating the possible charge transfers under 370 and 500 nm excitation.

The carrier dynamics can be analyzed also from the perspective of the CsPbI$_3$ domain. Within this context, **Figure 7c** presents the TA spectra at a 1 ps delay for the CsPbI$_3$ NCs, CsPbI$_3$-Pb$_4$S$_3$Cl$_2$, and CsPbI$_3$-CsPbCl$_3$-Pb$_4$S$_3$Cl$_2$ heterostructures. The GSB of CsPbI$_3$ excitons was observed at a shorter wavelength for CsPbI$_3$-Pb$_4$S$_3$Cl$_2$ (~670 nm) and CsPbI$_3$-CsPbCl$_3$-Pb$_4$S$_3$Cl$_2$ (~650 nm) compared to the isolated CsPbI$_3$ NCs (680 nm), in line with the steady-state PL and absorption data reported in **Figures 1a and S10.** This suggests possible differences in NC size and/or the presence of Cl$^-$ ions in the CsPbI$_3$ domain. Concurrently, bleach signals at ~620 nm were retrieved in the partially exchanged sample, which in turn could reflect the presence of alloying or populations of exchanged nano-heterostructures at different stages of exchange within the sample. The observation of GSB of CsPbI$_3$ (**Figure 7c**) and CsPbCl$_3$ (**Figure 7a**) excitons, together with the fast decay of CsPbCl$_3$ exciton GSB in the partially exchanged heterostructures are another indication of the presence of a $CsPbCl_3$-$CsPbI_3$ interface, which further supports the electron microscopy data. In addition, the partially exchanged heterostructures exhibit a slower rise time of the $CsPbI_3$ exciton GSB compared to the fully exchanged heterostructure under 370 nm pump illumination (**Figure S11a**). This is in line with a charge transfer from the $CsPbCl_3$ domain to the $CsPbI_3$ domain.[25] Conversely, when the pump excitation is below the $CsPbCl_3$ band gap (i.e. 500 nm), and only the $CsPbI_3$ domain can be excited, the rise profile of the $CsPbI_3$ exciton GSB is similar in both partially and fully exchanged heterostructures (**Figure S11b**). The decay of $CsPbI_3$ GSB (**Figure S12a and S12b**) is faster in the case of fully exchanged sample compared to both partially exchanged heterostructure and free-standing $CsPbI_3$ NCs. Additional processes at the newly formed $CsPbI_3$-$Pb_4S_3Cl_2$ interface are indeed expected from the band structure calculations in fully exchanged heterostructure, where carrier trapping due to defects could result in a faster decay of $CsPbI_3$ exciton GSB (**Figure 6c** and **Figure 7d**). In the case of partially exchanged sample, the presence of the $CsPbCl_3$ domain hinders any such process, and this is also supported by the calculations presented above. **Figure 7d** provides a schematic energy diagram illustrating the system: the wide-bandgap $CsPbCl_3$ domain (light blue) is centrally located, and, on either side, there are narrower bandgap $Pb_4S_3Cl_2$ (grey) and $CsPbI_3$ (light red) domains. Upon photoexcitation, charge transfer occurs preferentially and more rapidly to the $CsPbI_3$ domain, highlighting its efficient charge transfer properties. The $CsPbCl_3$ domain also acts as an energy barrier between the $Pb_4S_3Cl_2$ and $CsPbI_3$ domains, facilitating the radiative recombination in the $CsPbI_3$ region.

Finally, time-dependent stability tests were performed on the fully exchanged $CsPbI_3$–$Pb_4S_3Cl_2$ heterostructures. The results are reported in Figure S13 of the Supporting Information. The heterodimer morphology was preserved for the first eight days, with no evidence of secondary phases in either optical absorption or photoluminescence spectra. At longer times, however, heterodimers were no longer observed and mostly isolated particles were present. This reduced stability is consistent with the large lattice mismatch between $CsPbI_3$ and $Pb_4S_3Cl_2$, which limits the long-term preservation of a stable epitaxial interface

## 3. Conclusion

This study demonstrates that $CsPbCl_3$-$Pb_4S_3Cl_2$ NC heterostructures enable controlled Cl→I halide exchange, unlike the rapid, limited alloying in free-standing $CsPbCl_3$ NCs. The epitaxial perovskite-chalcohalide interface directs a gradual exchange, with iodide ions substituting chloride ions in the perovskite domain farthest from the interface, forming segmented $CsPbI_3$-$CsPbCl_3$-$Pb_4S_3Cl_2$ heterostructures with a unique $CsPbI_3$-$CsPbCl_3$ epitaxial interface. DFT calculations confirm that this configuration minimizes the total energy of the system, and it is further stabilized by iodine incorporation at the perovskite-chalcohalide interface. As the exchange progresses, the $CsPbCl_3$ domain shrinks to a few unit cells in projection and exhibiting iodine alloying, with $I^-$ ions preferentially aligned perpendicular to the $CsPbCl_3$-$CsPbI_3$ interfaces. These configurations also correspond to energetically favorable states, as supported by computational modeling. Optical analyses reveal type-I band alignments in intermediate heterostructures (both at the $CsPbCl_3$-$CsPbI_3$ and $CsPbCl_3$-$Pb_4S_3Cl_2$ interfaces). Especially for the $CsPbCl_3$-$CsPbI_3$ interface, this band alignment enables efficient carrier transfer to the $CsPbI_3$ domain, resulting in red emission.

In conclusion, the $CsPbCl_3$-$Pb_4S_3Cl_2$ epitaxial heterostructures strongly influence the Cl→I exchange mechanism, compared to free-standing $CsPbCl_3$ NCs, leading to novel segmented heterostructures sustaining a $CsPbI_3$-$CsPbCl_3$ interface, despite significant lattice mismatch between these two halide perovskite lattices. This behavior positions such heterostructures as a promising platform to induce controlled phase segregation and even create heterostructures with substantial lattice mismatch that are otherwise impossible to achieve via direct synthesis or by post-synthesis anion exchange in free-standing NCs. Such tunability in phase and band alignment can be further exploited in light-emitting applications, while engineering a type-II alignment promoting charge separation could specifically benefit photocatalysis or photovoltaics, offering a versatile platform for advanced optoelectronic and photocatalytic materials.

## 4. Experimental Section/Methods

*Materials*. Cesium carbonate ($Cs_2CO_3$, 99,9%), lead chloride ($PbCl_2$, >98%), lead iodide ($PbI_2$, >98%), lead acetate trihydrate ($Pb(CH_3COO)_2·3H_2O$, 99.99%), dodecanethiol (DDT, 99.9%), elemental sulfur (S, >99%), 1-octadecene (ODE, $C_{18}H_{36}$, 90%), oleic acid (OA, $C_{18}H_{34}O_2$, 90%), oleylamine (OLAm, $C_{18}H_{37}N$, 70%), anhydrous toluene ($C_7H_8$, >99.8%), ethyl acetate ($C_4H_8O_2$), fumed silica powder ($SiO_2$), and hexane (99.8%) were purchased from Merck. All chemicals were used without further purification.

*$PbCl_2$ stock solution*. $PbCl_2$ (2 mmol), 30 mL of octadecene, 5 mL of oleylamine, and 5 mL of oleic acid were loaded in a 100 mL 3-neck flask. The mixture was first degassed for 30 minutes at room temperature and then for 30 minutes at 110 °C under stirring. Finally, the solution was heated up to 150 °C under $N_2$ until the salt was completely dissolved. The resulting solution was transferred into a N$_2$-filled glass vial and was stored inside the glovebox for further use.

*Cs-OL stock solution*. Cs$_2$CO$_3$ (1 mmol), 2.5 mL of oleic acid, and 8.75 mL of octadecene were loaded in a 25 mL 3-neck flask. The mixture was first degassed for 30 minutes at room temperature and then for 1 hour at 110 °C under stirring. The resulting solution was transferred into a N$_2$-filled glass vial and was stored inside the glovebox for further use.

*Pb-OL stock solution*. 1 mmol of lead acetate trihydrate, 650 µL of oleic acid, and 9.35 mL of octadecene were loaded in a 50 mL 3-neck flask. The mixture was first degassed for 30 minutes at room temperature and then for 1 hour at 110 °C under stirring. The resulting solution was transferred into a N$_2$-filled glass vial and was stored inside the glovebox for further use.

*S-ODE precursor solution*. 0.5 mmol of sulfur powder and 5 mL ODE (previously degassed for 1 hour at 110 °C) were loaded in a 7 mL vial inside the glovebox. The resulting mixture was sonicated until the sulfur powder was completely dissolved.

*PbI$_2$ stock solution.* 1 mmol of PbBr$_2$, 2.5 mL of oleic acid, 2.5 mL of oleylamine, and 15 mL of octadecene were loaded in a 40 mL vial. The mixture was degassed for 30 minutes at room temperature and then for 30 minutes at 110 °C. Then, the solution was heated up to 150 °C under N$_2$ until the salt was completely dissolved. The resulting solution was transferred into a N$_2$-filled glass vial and was stored inside a glovebox for further use.

*Oleylammonium Iodide precursor solution.* I$_2$ powder (6 mmol), oleylamine (0.027 mol), and ODE (21 mL) in a 40 mL glass vial. The vial was placed on the hot plate and degassed under vacuum for 30 minutes at 110 °C. The resulting solution was cooled down at room temperature and stored under nitrogen.

*CsPbCl$_3$ clusters synthesis.* CsPbCl$_3$ clusters were synthesized following a previously reported method with slight modifications.[14] Briefly, 4 mL of the PbCl$_2$ stock solution was transferred into a 20 mL vial. The solution was heated up to 50 °C and then 0.25 mL of Cs-OL stock solution was injected into the PbCl$_2$ stock solution. The resulting mixture was kept under stirring at 50 °C for 30 min. The resulting solution was purified by centrifugation (8000 rpm, 5 min), and the supernatant was discarded. Finally, the precipitate was redispersed in a solution containing 0.9 mL of previously degassed octadecene.

*CsPbCl$_3$-Pb$_4$S$_3$Cl$_2$ NC heterostructures synthesis.* In a typical synthesis, 4.0 mL of previously degassed octadecene was loaded into a 20 mL N$_2$-filled vial. Then, the vial was heated up at 200 °C for 5 min. After this time, the following precursor solutions were injected into octadecene: 0.1 mL of Pb-OL solution, 0.2 mL of DDT-ODE solution (500 uL DDT and 4.5 mL of previously degassed octadecene), clusters solution (mentioned above), and 0.1 mL of sulfur source (S-ODE). The reaction was kept under stirring for the corresponding reaction time of 5 min and finally quenched in an ice-water bath. Purification steps included the addition of ethyl acetate (3 mL) into the crude solution and centrifugation (6000 rpm, 5 min). Afterward, the supernatant was discarded, and the precipitate was redispersed in 2 mL of toluene. To assess the concentration, the optical absorption spectrum of the NC solution was measured after diluting 25 μL of the NC solution to 2.5 mL toluene, resulting in an optical density of 0.14 at 370 nm.

*Synthesis of CsPbCl$_3$ NCs.* CsPbCl$_3$ colloidal NCs, were synthesized by injecting the previously made CsPbCl$_3$ clusters (dispersed in 0.9 mL ODE) in 4 mL pre-degassed ODE at 160 °C. The reaction was quenched by an ice/water bath after 10 minutes. Then, 2.5 mL of anhydrous methyl acetate was added to the crude solution, and the solution was centrifuged for 5 minutes at 6000 rpm. Finally, the precipitate was redispersed in toluene

*Synthesis of CsPbI$_3$ NCs.* CsPbI$_3$ colloidal NCs, were synthesized following the previously reported process.[26] Cesium carbonate (16 mg), lead acetate trihydrate (76 mg), OA (0.2 ml) and octadecene (5 ml) were added in a 40 ml glass vial. The solution dried for 1 h at 100 °C. Then, the temperature was adjusted at 165 °C, and 2 ml of previously made oleylammonium iodide solution was swiftly injected under nitrogen. The reaction was quenched by an ice/water bath after 10 seconds. Then, 5 mL of anhydrous methyl acetate was added to the crude solution, and the solution was centrifuged for 5 minutes at 6000 rpm. Finally, the precipitate was redispersed in toluene.

*Cl→I exchange reactions.* Halide exchange reactions were performed in a glovebox. In a typical reaction, 1 mL of the CsPbCl$_3$-Pb$_4$S$_3$Cl$_2$ solution was used for each reaction, and different amounts of PbI$_2$ solution (ranging from 20 μL to 300 uL) were added under stirring at room temperature. When the reaction reached equilibrium, 600 uL of ethyl acetate was added to the solution. Finally, the NCs were collected by centrifugation at 6000 rpm for 5 minutes and redispersed in anhydrous toluene for further characterization.

*Optical characterization.* Optical absorption spectra were collected on a Varian Cary 300 UV-VIS absorption spectrophotometer, while PL spectra were recorded by a Varian Cary Eclipse spectrophotometer using an excitation wavelength of 350 nm for the pristine sample and 500 nm for the exchanged samples. The NCs solutions were diluted in toluene in quartz cuvettes (path length = 1 cm) to a maximum optical density below 1.0.

*X-ray Powder Diffraction.* X-ray powder diffraction measurements were performed on a PANanalytical Empyrean X-ray diffractometer, equipped with a 1.8 kW Cu Kα ceramic anode and a PIXcel3D 2 × 2 area detector, operating at 45 kV and 40 mA. NCs solutions were mixed with fumed silica and dried to minimize the preferential orientation. Finally, the powder was placed on a zero-diffraction silicon substrate to perform the measurements.

*Scanning Transmission electron microscopy (STEM).* High-resolution scanning TEM (HRSTEM) images were acquired on a probe-corrected ThermoFisher Spectra 30-300 S/TEM operated at 300 kV, using a HAADF detector with a beam current of a few tens of pA to limit beam damage to the electron beam sensitive perovskite nanocrystals. The convergence angle was set to 25 mrad, corresponding to a sub-angstrom electron beam. The calculated electron dose per frame was between 240 to 9500 e/A$^2$ for HAADF-STEM imaging. Compositional maps were acquired using Velox software, with a probe current of ~150 pA and rapid rastered scanning. To improve the image quality of the elemental maps, a built-in Gaussian blur filter with a sigma value of 1.0 pixels was applied. The energy-dispersive X-ray (EDX) spectroscopy signal was acquired on a Dual-X setup comprising two detectors on either side of the sample, for a total acquisition solid angle of 1.76 Sr. The lattice parameters for HRSTEM images were mapped based on Pb atomic columns, positions of which were fitted using Gaussian functions.[27]

*Computational methodology.* To shed light onto the structural and electronic properties of the NC heterostructures, we carried out atomistic simulations with a density functional theory (DFT) Hamiltonian. The Perdew–Burke–Ernzerhof (PBE) exchange-correlation functional was employed.[28] Scalar-relativistic Goedecker–Teter–Hutter (GTH) pseudopotentials were used for all atoms, replacing the core electrons, while the valence electrons were described with a double-ζ basis set plus polarization functions[29] as implemented in the CP2K 2024.1 package.[30] The auxiliary plane–wave cutoff was set to 400 Ry. All calculations were performed in vacuum within a non-periodic simulation box of 52 × 68 × 52 Å. Structural relaxations were performed until the maximum residual force on atoms was below 4.5 × 10$^{-4}$ Ha/Bohr.

*4D-STEM characterization.* 4D-STEM imaging was performed with a probe-corrected Thermo Fisher Titan Themis at an acceleration voltage of 200 kV and a semi-convergence angle of the electron beam of 14 mrad. The NCs were deposited on a homemade graphene TEM-grid.[31] The 4D-STEM datasets were acquired with a custom-made Timepix3 detector2, which is an event-driven hybrid pixelated direct electron detector, in event-based format with 2048x2048 dimensions using a dwell time of 1 μs and a scanning step sizes of 0.2-0.1 Å. Final images were retrieved from 4D-STEM datasets using recently developed CNN approach.[20] This method allows recovering the phase and the amplitude components of the exit electron wave based on a 3x3 kernel of adjacent convergent beam electron diffraction patterns. Real space was binned two times, and the detector was binned to 64x64 for the reconstruction. The data generation code is available open source under https://github.com/ThFriedrich/airpi. The electron doses were estimated by determining the number of detected events on the detector and by considering the cluster size. The cluster size correlates with the number of pixels which are excited by a single electron and was determined to be 2.3 for 200 kV.[19] The octahedral tilt measurements were performed based on fitted Pb and X (halide) atomic columns with Gaussians by assessing the averaged angle between the lattice vectors and the vectors corresponding to the direction of PbX$_6$ octahedra, thus, reflecting Pb-Pb-X angle.[27] X-X-Cs angles analysis between two vectors was performed after fitting all atomic column positions with Gaussians and extracting the coordinates of required atomic columns (Cs and X).

*Transient Absorption Spectroscopy*. A tunable Ytterbium-based laser system (Pharos-SP-HP equipped with an optical parametric amplifier Orpheus from Light Conversion) was employed to perform the ultrafast transient absorption spectroscopy. The pulse duration of the pump was ≈160 fs whereas the instant response function, determined by the cross-correlation between pump and probe, was experimentally assessed to be around 230 fs. An excitation pump power of ≈2 mW was used. All the TA spectra have been acquired through the Harpia-TA (Light Conversion) system, where the supercontinuum white light is generated by focusing a portion of the second harmonic of the fundamental laser beam (at ≈515 nm) onto a sapphire crystal and the delay between the pump and probe pulses has been tuned by changing the length of the probe optical path. The experiments were performed at 50 kHz while the spot size of the pump/probe at the sample was measured to be approximately 120 μm and 60 μm, respectively. Data from the ultrafast experiments were corrected for chirp dispersion using a commercial software (CarpetView, Light Conversion).

**Acknowledgements**

Y.P.I. and G.D. acknowledge funding from the Italian Space Agency (Agenzia Spaziale Italiana, ASI) in the framework of the Research Day "Giornate della Ricerca Spaziale" initiative through the contract ASI N. 2023−4-U.0. N.L. and L.M. acknowledge funding from European Research Council through the ERC Advanced Grant NEHA (grant agreement n. 101095974). J.Z. and L.M. acknowledge funding from Ministero dell'Ambiente e della Sicurezza Energetica through the Project IEMAP (Italian Energy Materials Acceleration Platform) within the Italian Research Program ENEA-MASE (2021−2024 "Mission Innovation" (agreement 21A033302 GU no. 133/5−6−2021) as well as the extensive use of the CRESCO/ENEAGRID High Performance Computing infrastructure and the support of its management team.[32,33] I.S. S.V.A., S.B. and L.M. acknowledge financial support from the Research Foundation-Flanders (FWO) through a doctoral fellowship (FWO-SB Grant No. 1SHA024N) and project funding (G0A7723N). A.T. acknowledges support by the European Research Council under the European Union's Horizon 2020 Research and Innovation Program through the ERC Consolidator Grant REPLY (grant agreement n. 101002422). S. B. and L. M. acknowledge funding from the European Union's Horizon Europe grant MSCA SE DELIGHT (GA n. 101131111). We thank Ye Wu for insightful discussions on heterostructure synthesis.

**Data Availability Statement**

The data that support the findings of this study are available from the corresponding authors upon reasonable request.

**Perovskite-Chalcohalide Nanocrystal Heterostructures as a Platform for the Synthesis and Investigation of CsPbCl$_3$-CsPbI$_3$ Epitaxial Interface**

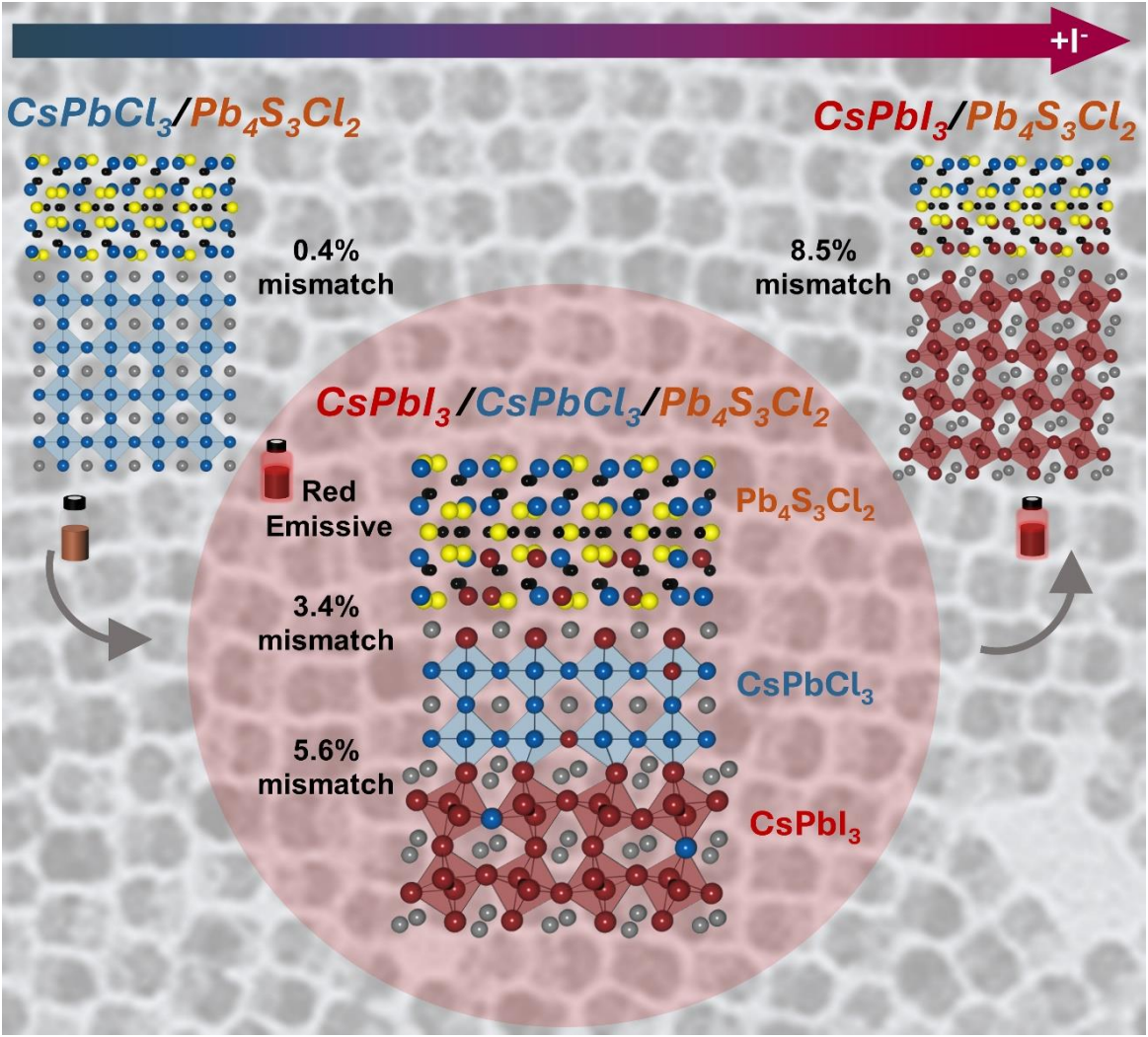

Supporting Information

**Halide Perovskite-Chalcohalide Nanocrystal Heterostructures as a Platform for the Synthesis and Investigation of the CsPbCl₃-CsPbI₃ Epitaxial Interface**

*Nikolaos Livakas, Irina Skvortsova, Juliette Zito, Yurii P. Ivanov, Aswin Asaithambi, Andrea Toma, Annick De Backer, Muhammad Imran, Sandra Van Aert, Giorgio Divitini, Ivan Infante\*, Sara Bals\* and Liberato Manna\**

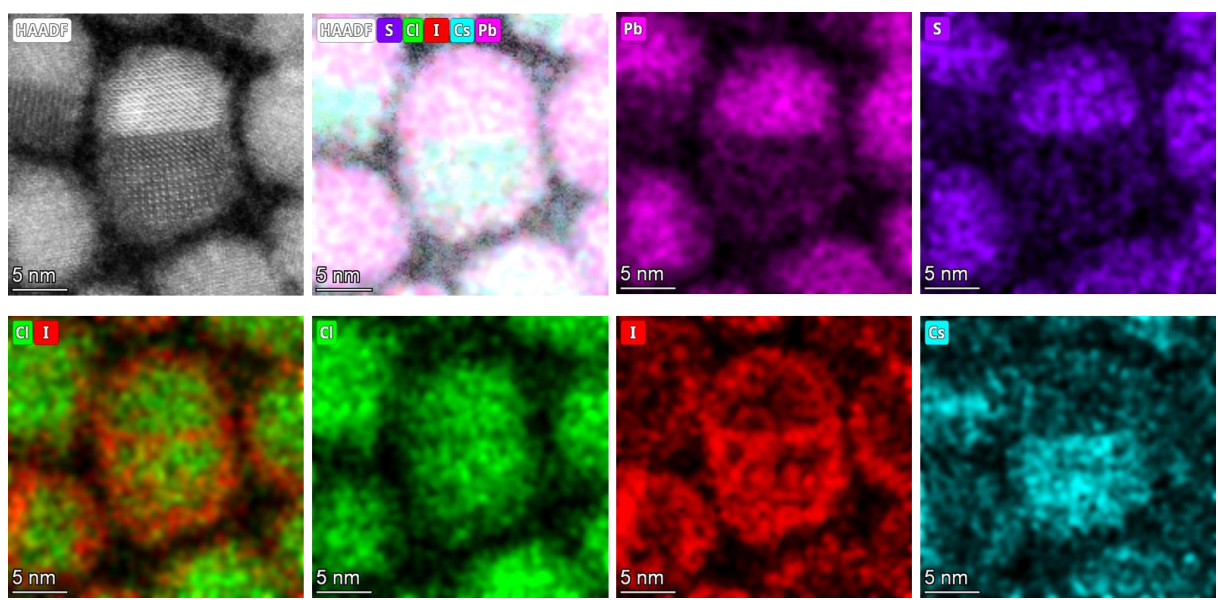

**Figure S1.** HAADF-STEM image of the partially exchanged segmented $CsPbI_3$-$CsPbCl_3$-$Pb_4S_3Cl_2$ heterostructure with the corresponding STEM-EDX elemental maps.

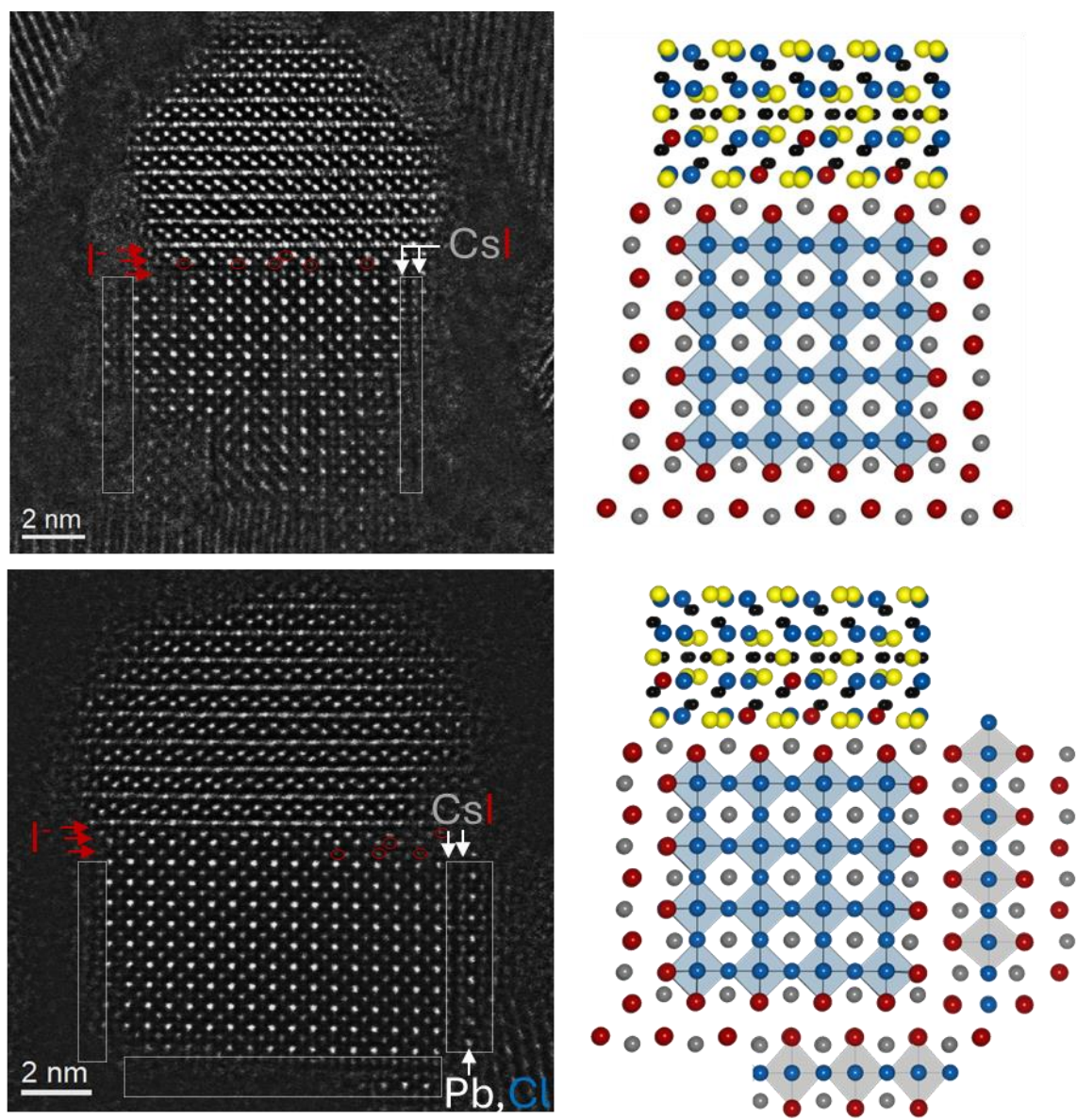

**Figure S2**. HAADF-STEM images of two nanocrystals captured at the early stages of the Cl→I exchange in CsPbCl$_3$–Pb$_4$S$_3$Cl$_2$ heterostructures. The early stage of the reaction may involve the formation of a CsI passivation layer on the CsPbCl$_3$ surface, accompanied by iodine incorporation at the perovskite–chalcohalide interface. The CsI surface layer, together with the underlying PbCl$_2$ layer of the CsPbCl$_3$ core, can be regarded as part of a Cs$_2$PbCl$_2$I$_2$ Ruddlesden–Popper (RP) phase, as in some cases nearly a complete RP unit cell is observed (bottom panel).

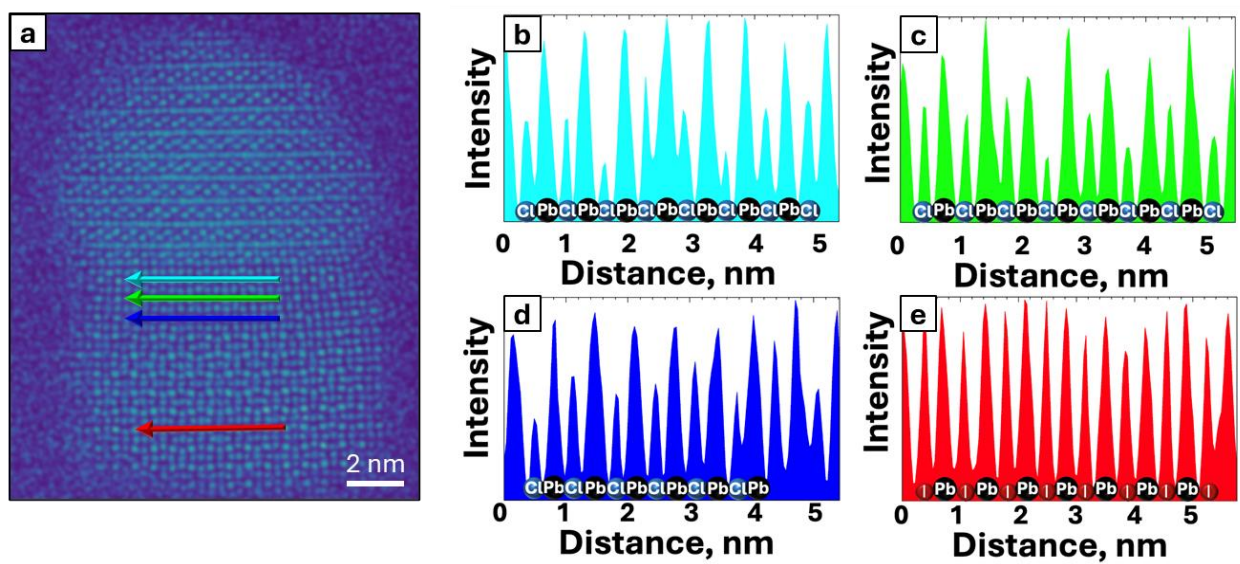

**Figure S3.** (a) Phase image obtained from CNN reconstruction of a 4D-STEM dataset of the partially exchanged $CsPbI_3$-$CsPbCl_3$-$Pb_4S_3Cl_2$ heterostructures presented in Figure 3b. The arrows correspond to the intensity profiles presented in panels (b-e) highlighting the different perovskite subdomains ($CsPbI_3$ at the bottom of the nanocrystal and $CsPbCl_3$ close to the chalcohalide domain).

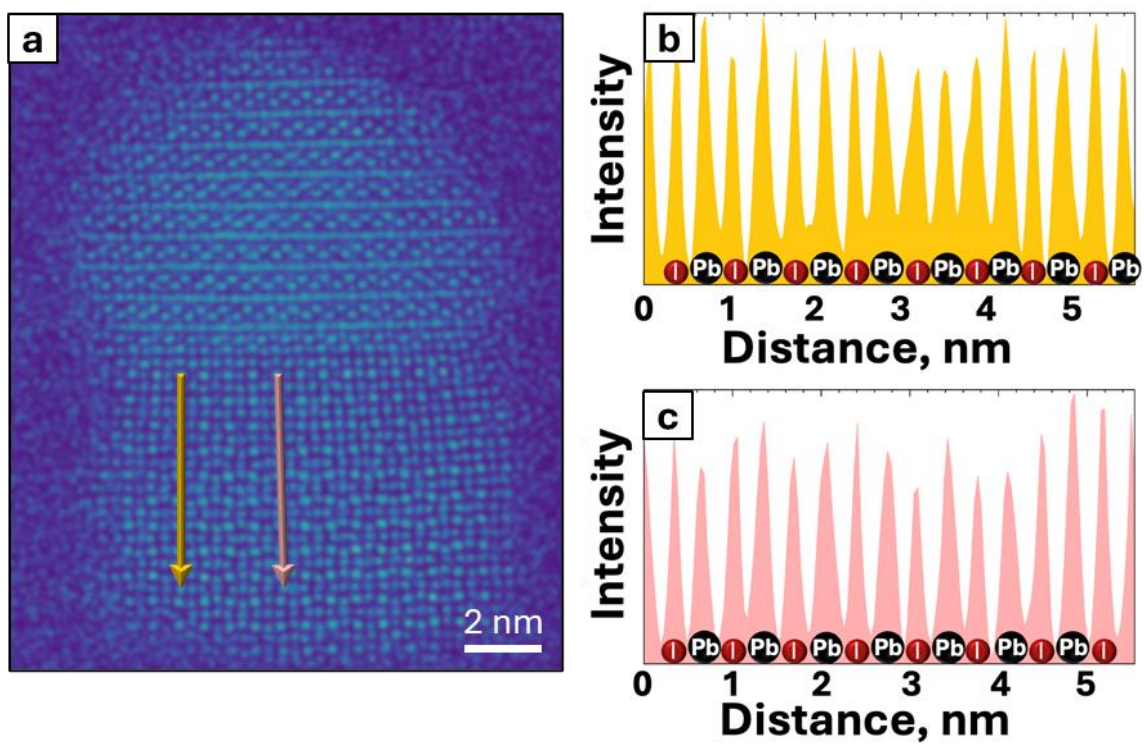

**Figure S4.** Phase image obtained from CNN reconstruction of a 4D-STEM dataset of the partially exchanged $CsPbI_3$-$CsPbCl_3$-$Pb_4S_3Cl_2$ heterostructures presented in Figure 3b. The arrows correspond to the intensity profiles presented in panels (b and c) highlighting that the $CsPbCl_3$ lattice is alloyed with iodine ions and those ions are concentrated in the axial direction.

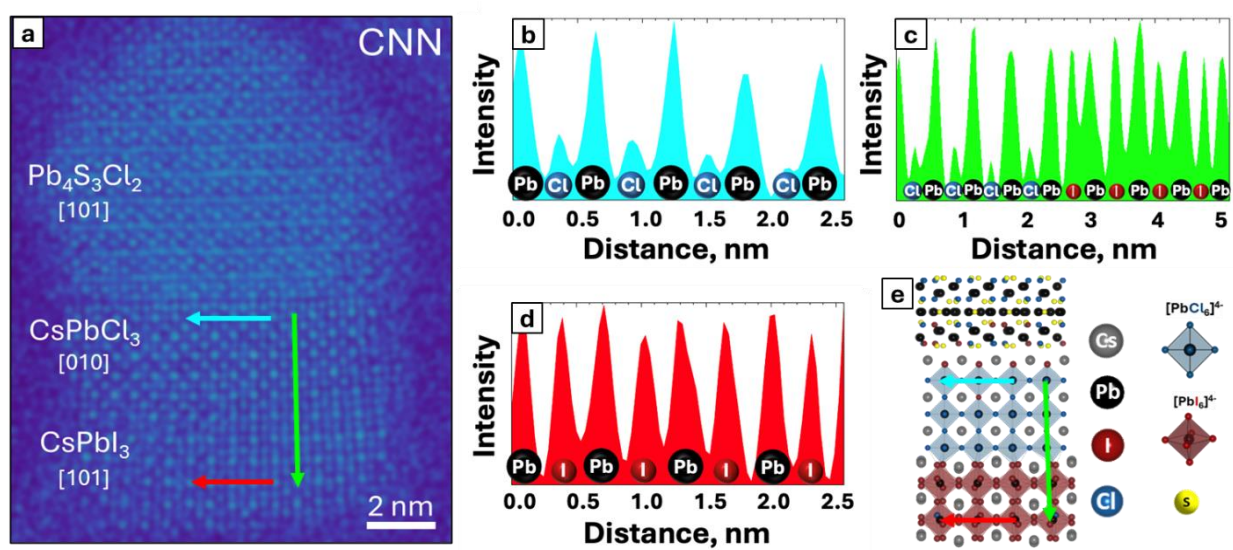

**Figure S5**. (a) Phase image obtained from CNN reconstruction of a 4D-STEM data set of the partially exchanged $CsPbI_3$-$CsPbCl_3$-$Pb_4S_3Cl_2$ heterostructures with $CsPbI_3$ aligned in [101]

zone axis. Gaussian-filter was applied to the image. (b,c,d) Intensity profiles correspond to the arrows presented in panel (a). (e) Crystallographic model showcasing the multidomain structure presented in panel (a). In contrast to the intermediate case presented in Figure 3b, here the CsPbCl$_3$ domain is bigger than few atomic rows, and there is no iodine alloying within its lattice.

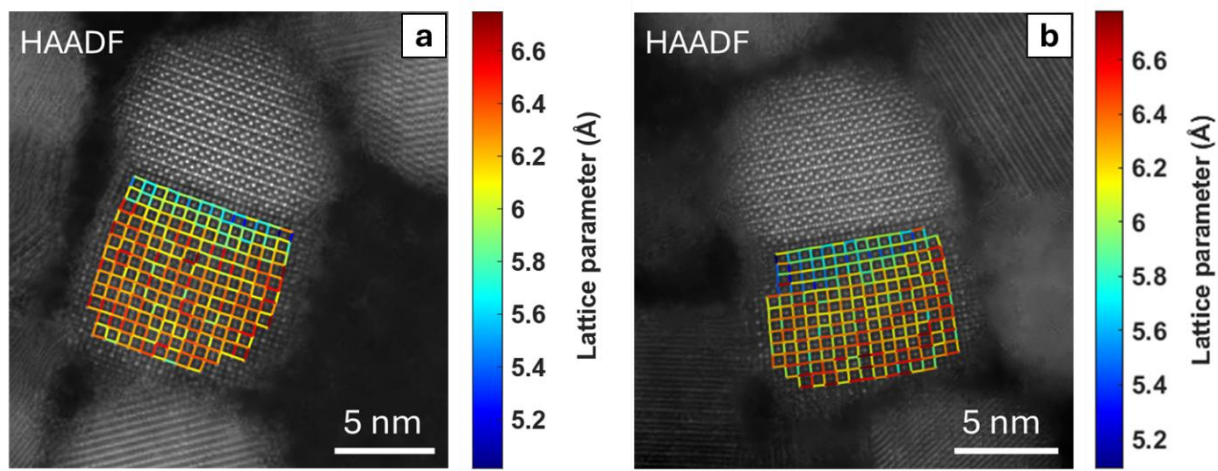

**Figure S6.** Lattice parameters quantified based on HAADF-STEM images of partially exchanged heterostructures, with (a) [010] oriented and (b) [101] oriented CsPbI$_3$ domains.

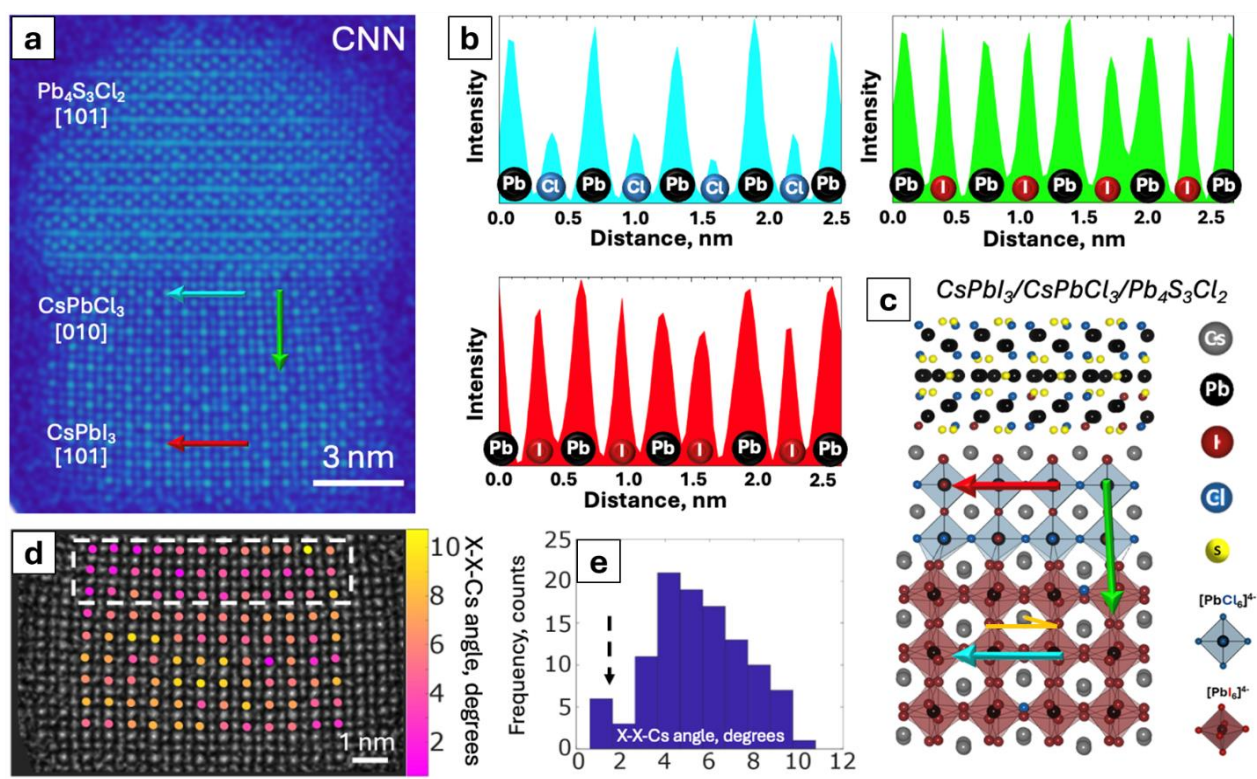

**Figure S7.** (a) Phase image obtained from CNN reconstruction of a 4D-STEM dataset of the partially exchanged $CsPbI_3$-$CsPbCl_3$-$Pb_4S_3Cl_2$ heterostructures with $CsPbI_3$ in [101] zone axis. Gaussian-filter was applied to the image. (b) Intensity profiles correspond to the arrows presented in panel (a). (c) Crystallographic model corresponding to the multidomain structure presented in panel (a). (d) Quantification of the X-X-Cs (X: halides) angles represented by a map of the angles throughout the perovskite domain for the partially exchanged sample. (e) Histogram of the measured X-X-Cs angles.

To quantify structural differences for the partially exchanged heterostructures with $CsPbI_3$ in [101] orientation, we used X-X-Cs (where X corresponds to halides) angles as key indicators rather than measuring octahedral tilt (Pb-Pb-X angles) since the tilt cannot be not directly observed along this projection. The X-X-Cs values were compared for the intermediate $CsPbCl_3$-$CsPbI_3$-$Pb_4S_3Cl_2$ (**Figure S7d,e**) and fully exchanged $CsPbI_3$-$Pb_4S_3Cl_2$ heterostructures (**Figure S8**). For the cubic $CsPbCl_3$ lattice, which is expected to be present close to the interface with the chalcohalide domain, the angles are close to 0 degrees (**Figure S7e, black dashed arrow**). Again, when moving away from the interface the X-X-Cs angles increase up to 6-7 degrees. These values are in close agreement with those observed for the fully exchanged $CsPbI_3$ (**Figure S8**). These observations therefore confirm the structural segmentation for the partially exchanged heterostructures.

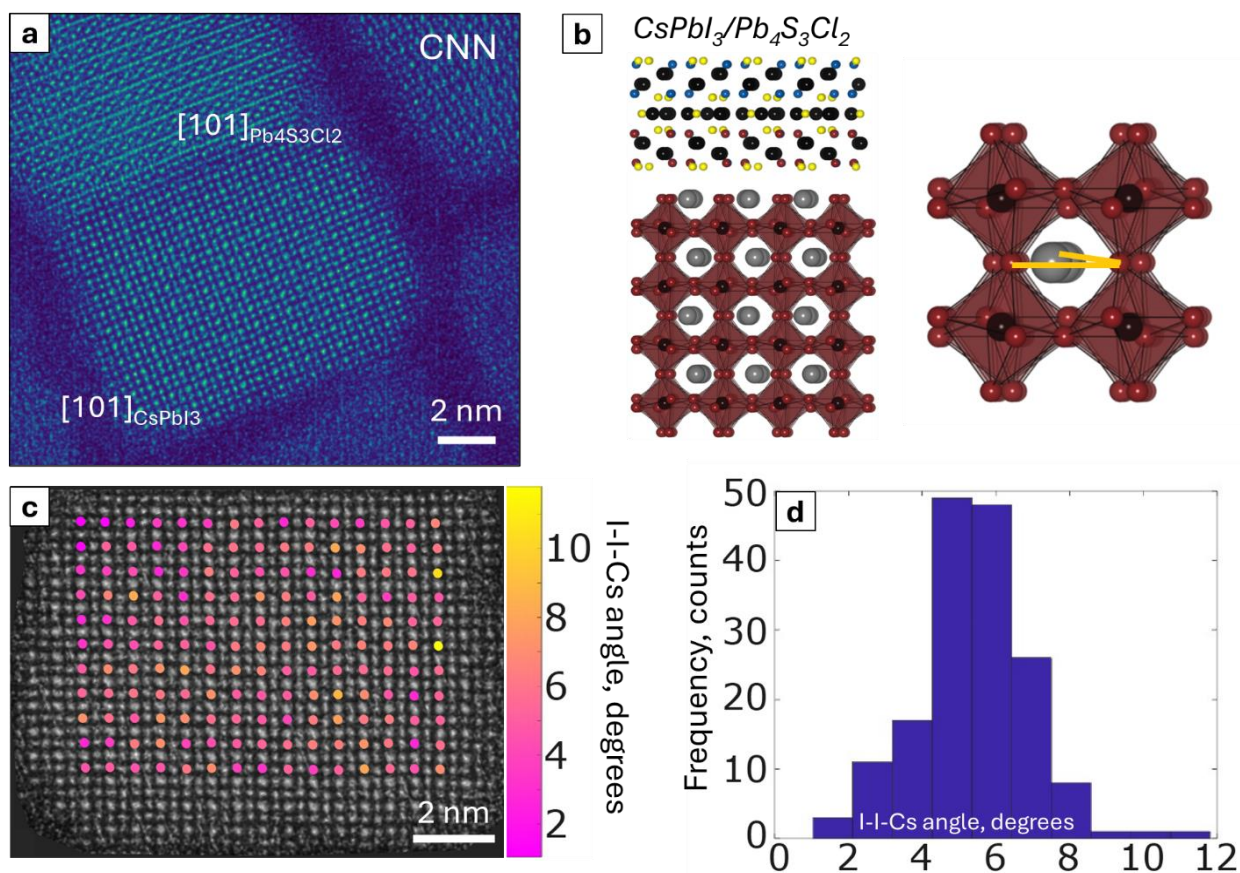

**Figure S8.** (a) Phase image obtained from CNN reconstruction of a 4D-STEM data set of the fully exchanged $CsPbI_3$-$Pb_4S_3Cl_2$ heterostructures with $CsPbI_3$ in the [101] zone axis. Gaussian-filter was applied to the image. (b) Structural model corresponding to the image presented in panel (a). (c) Quantification of the I-I-Cs angles in the equatorial direction is represented by a map of the angles throughout the perovskite domain. (d) Histogram of the measured I-I-Cs angles.

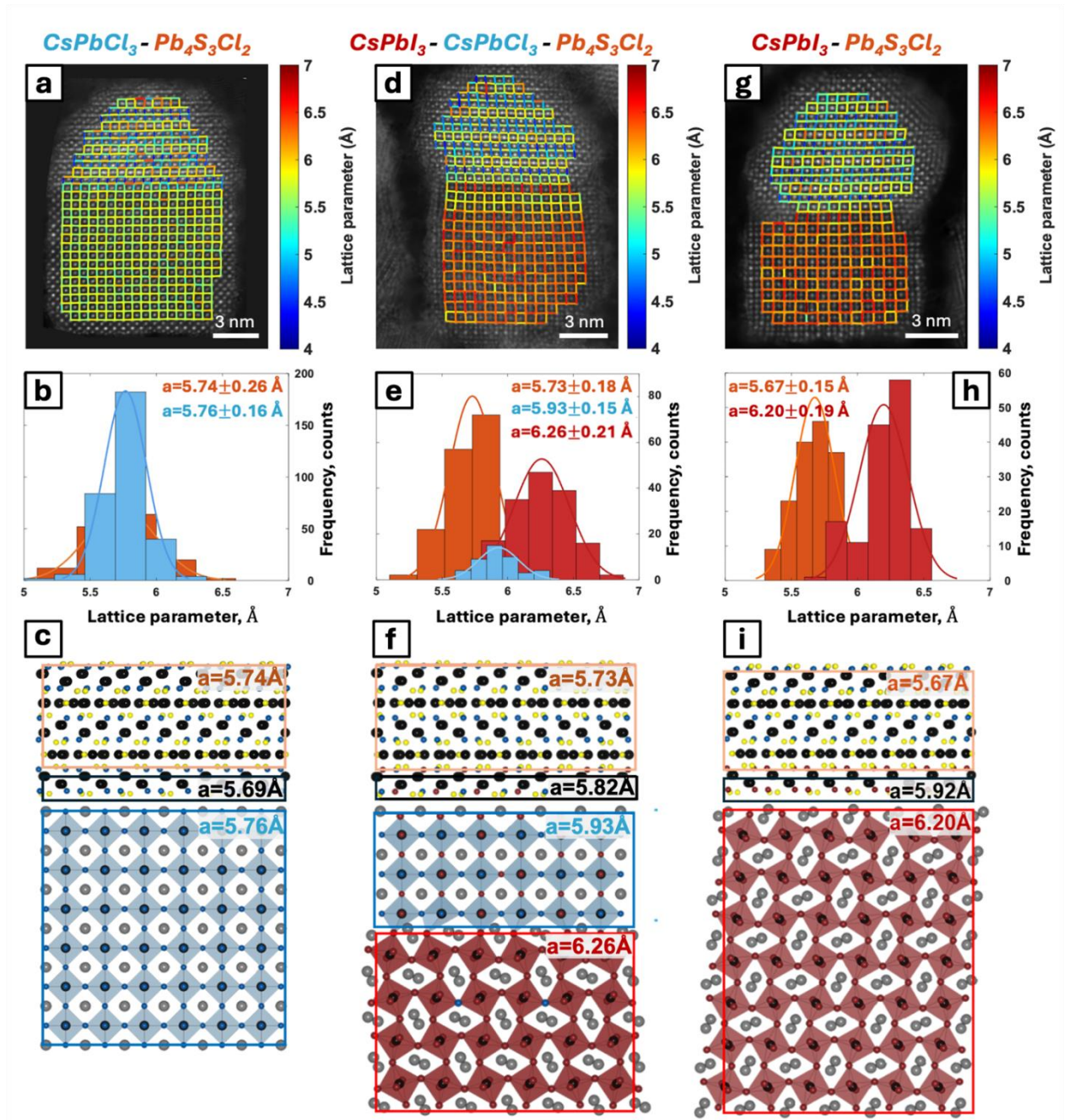

**Figure S9.** Lattice parameters analysis performed on HAADF-STEM images with corresponding distributions of the lattice parameters parallel to the perovskite-chalcohalide interface and structural models for (a,b,c) pristine, (d,e,f) partially and (g,h,i) fully exchanged heterostructures. Orange histograms and Gaussians belong to chalcohalide phase, blue histograms and curves correspond to $CsPbCl_3$ part and red ones belong to $CsPbI_3$ domain. For (e), the first three Pb-containing atomic rows in perovskite region (close to chalcohalide domain) are considered as $CsPb(Cl_xI_{1-x})_3$ and the rest is [010]-oriented $CsPbI_3$. In (g) $CsPbI_3$ domain is also oriented along the [010] direction. The average distances at the perovskite-chalcohalide interface are 5.69±0.08 Å for (a), 5.82±0.14 Å for (d) and 5.92±0.18 Å for (g),

represented in black boxes in panels (c,f,i). This perovskite-chalcohalide interface corresponds to a monolayer with a $Pb_4S_3Cl_2$ composition for the pristine sample, and as $Pb_4S_3(Cl_{1-x}I_x)_2$ composition for the partially and fully exchanged cases.

**Table S1.** Reaction energies associated to the Cl→I anion exchange in five layers of the perovskite domain, either starting from the $CsPbCl_3$-$Pb_4S_3Cl_2$ interface or from the bottom of the perovskite domain. All energies are normalized to the number of exchanged halide ions. In the last column, the positive sign indicates that the exchange is always more favored from the bottom.

| N layers | $\Delta E_{exchange\ interface}$ (kcal/mol) | $\Delta E_{exchange\ bottom}$ (kcal/mol) | $\Delta E_{exchange\ interface}$ − $\Delta E_{exchange\ bottom}$ (kcal/mol) |
|---|---|---|---|
| 1 | 4.32 | 0.53 | 3.78 |
| 2 | 3.79 | 3.39 | 0.40 |
| 3 | 4.11 | 3.58 | 0.53 |
| 4 | 3.80 | 3.54 | 0.26 |
| 5 | 3.95 | 3.71 | 0.25 |

**Table S2.** Reaction energies associated to the partial Cl→I anion exchange in two interfacial layers of the chalcohalide domain. All energies are normalized to the number of exchanged halide ions. In the third column, the positive sign indicates that the initial exchange (I/Cl ratio from 0% to 20%) is more favored than the subsequent exchanges. For each Cl→I exchange in the chalcohalide domain, the corresponding variation of the interfacial strain in the perovskite subdomain is also reported. The initial exchange (I/Cl ratio from 0% to 20%) corresponds to the largest release of interfacial strain.

| Exchange step (I/Cl ratio, %) | $\Delta E_{exchange}$ (kcal/mol) | $\Delta E_{exchange} - \Delta E_{exchange\ (0\%\rightarrow 20\%)}$ (kcal/mol) | strain (%) |
|---|---|---|---|
| 0%→20% | 1.72 | 0.00 | 1.80%→1.58% |
| 20%→40% | 2.78 | 1.05 | 1.58%→1.44% |
| 40%→60% | 2.56 | 0.83 | 1.44%→1.45% |
| 60%→80% | 2.58 | 0.85 | 1.45%→1.40% |
| 80%→100% | 2.82 | 1.09 | 1.40%→1.51% |

**Table S3.** Reaction energies associated to the partial Cl→I anion exchange in the $CsPbCl_3$ domain. Four possible distributions of the I ions are reported: random in the $CsPbCl_3$ domain; ordered in a CsI layer perpendicular to the interface; ordered in a CsI layer parallel to the interface; ordered in a $PbI_2$ layer perpendicular to the interface. In the last column, the negative sign indicates that the ordered configurations are always preferred over the random configuration.

| I distribution | $\Delta E_{exchange}$ (kcal/mol) | $\Delta E_{exchange} - \Delta E_{exchange\ (random)}$ (kcal/mol) |
|---|---|---|
| random | 87.13 | 0 |
| CsI layer ⊥ | 83.91 | -3.22 |
| CsI layer ∥ | 86.95 | -0.18 |
| PbI2 layer ⊥ | 70.38 | -16.75 |

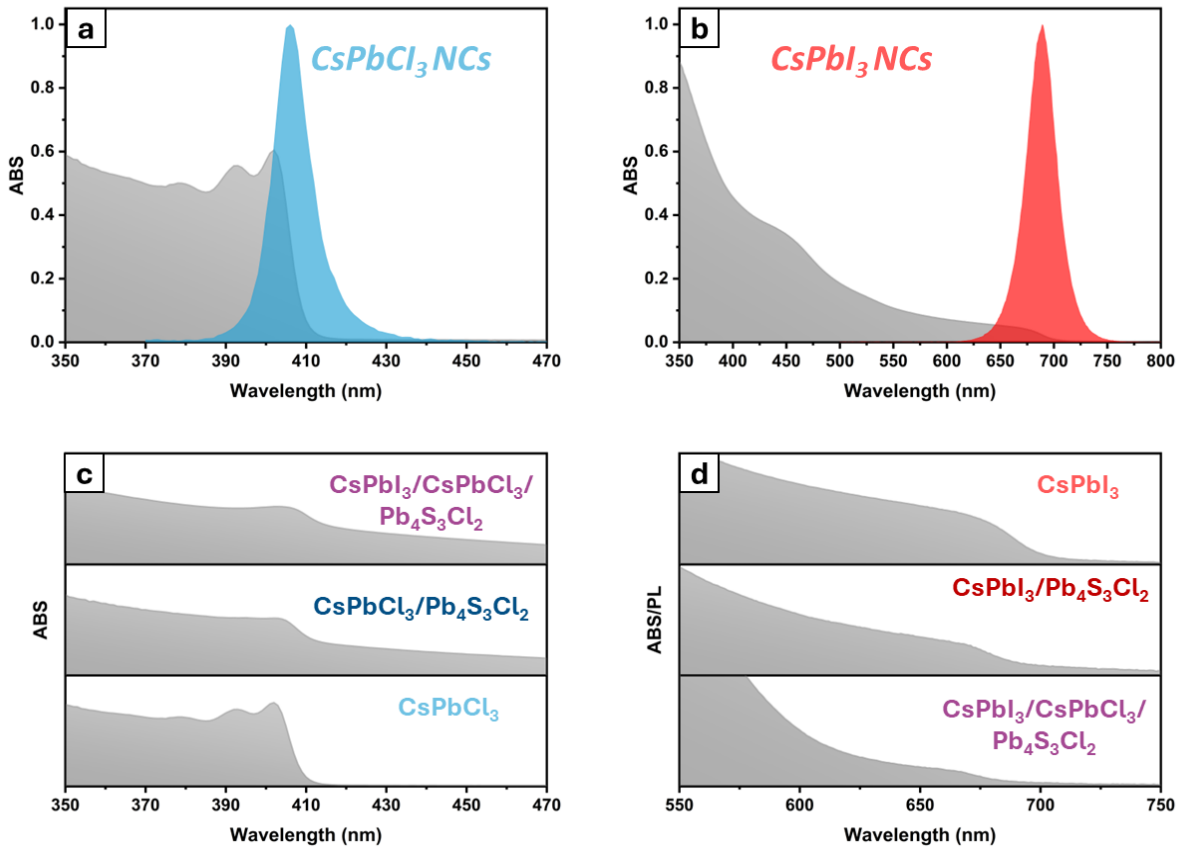

**Figure S10.** Optical absorption (grey) and PL (colored) spectra of as-synthesized pure (a) $CsPbCl_3$ and (b) $CsPbI_3$ NCs. (c) Comparison of the optical absorption spectra between free-standing $CsPbCl_3$ NC, $CsPbCl_3$-$Pb_4S_3Cl_2$, and partially exchanged $CsPbI_3$-$CsPbCl_3$-$Pb_4S_3Cl_2$ heterostructures. (d) Comparison of the optical absorption spectra between free standing $CsPbI_3$ NC, partially exchanged $CsPbI_3$-$CsPbCl_3$-$Pb_4S_3Cl_2$, and fully exchanged $CsPbI_3$-$Pb_4S_3Cl_2$ heterostructures.

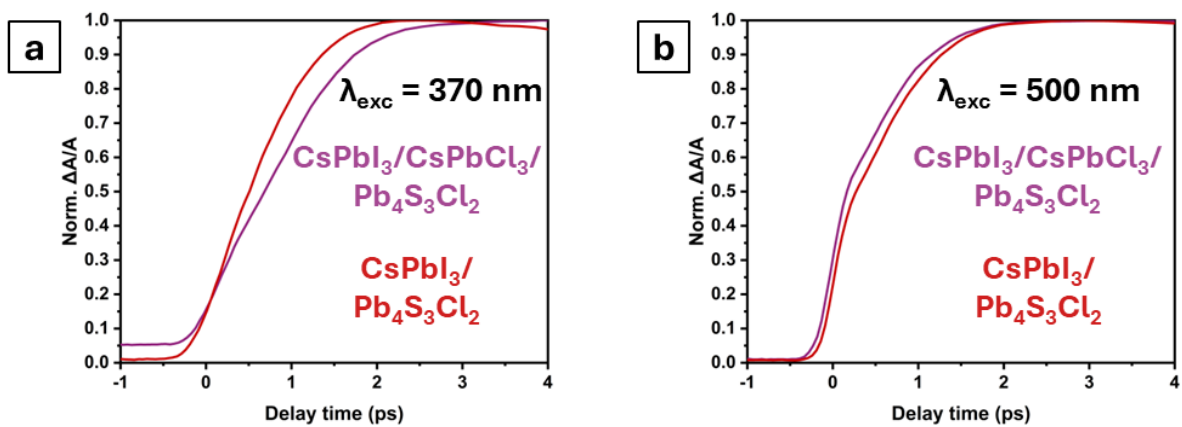

**Figure S11.** Transient absorption rise profile of the CsPbI$_3$ exciton (ground-state bleaching, GSB) under (a) 370 nm excitation and (b) 500 nm excitation.

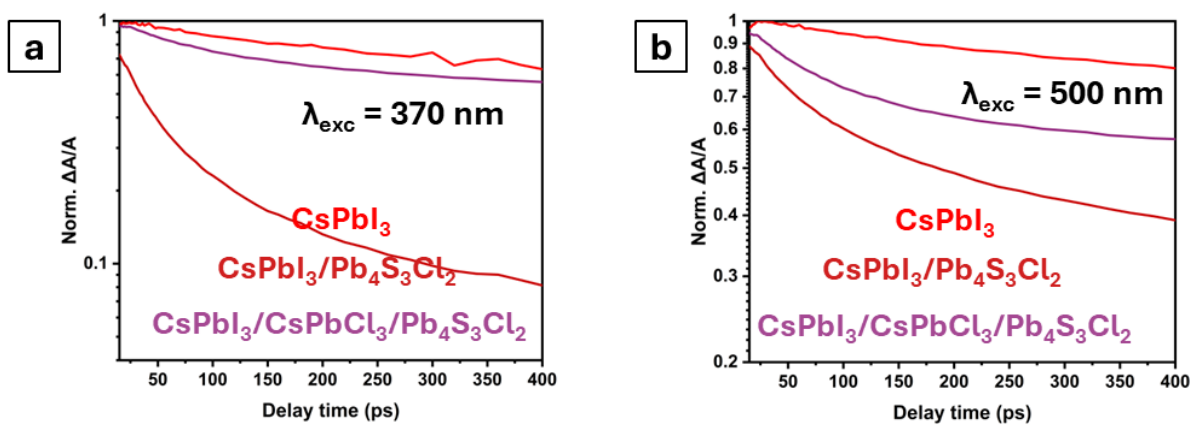

**Figure S12.** Transient absorption decay profile of the CsPbI$_3$ exciton GSB under (a) 370 nm excitation and (b) 500 nm excitation.

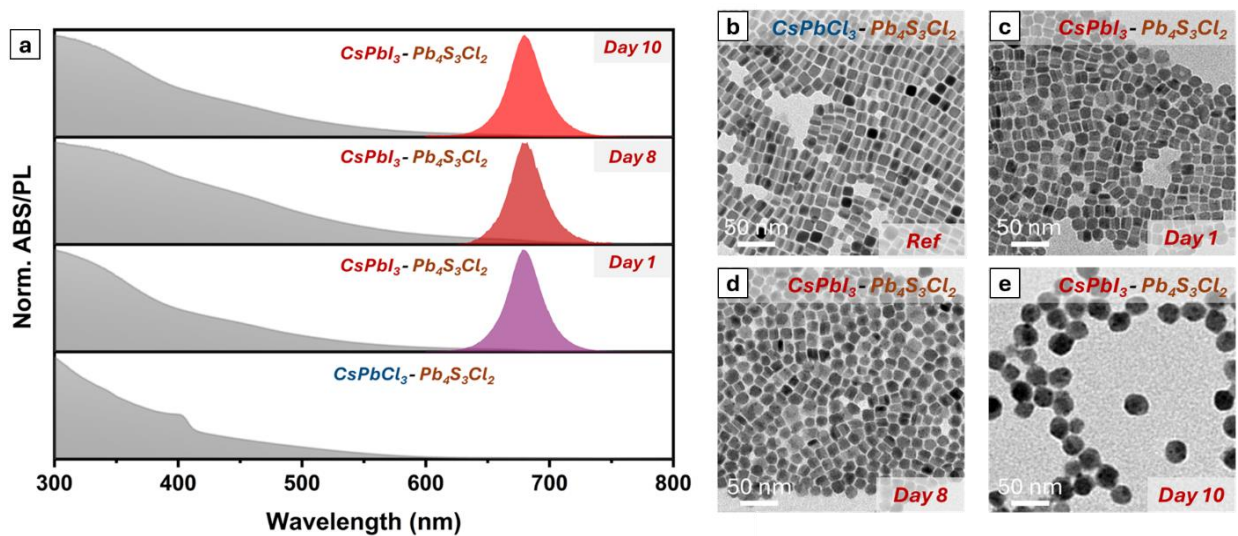

**Figure S13.** Steady-state optical and electron microscopy analyses of the initial $CsPbCl_3$–$Pb_4S_3Cl_2$ and fully exchanged $CsPbI_3$–$Pb_4S_3Cl_2$ NC heterostructures. (a) Optical absorption (grey) and PL (colored) spectra of the pristine and fully exchanged heterostructures measured after 1, 8, and 10 days of halide exchange reaction. TEM images of (b) pristine heterostructures and fully exchanged heterostructures after (c) 1 day, (d) 8 days, and (e) 10 days of reaction.